# Evidence for a Global Warming at the Termination I Boundary and Its Possible Cosmic Dust Cause

Paul A. LaViolette

The Starburst Foundation
6706 N. Chestnut Ave., #102
Fresno, CA 93710 USA

**Abstract**

A comparison of northern and southern hemispheric paleotemperature profiles suggests that the Bölling-Alleröd Interstadial, Younger Dryas stadial, and subsequent Preboreal warming which occurred at the end of the last ice age were characterized by temperatures that changed synchronously in various parts of the world, implying that these climatic oscillations were produced by significant changes in the Earth's energy balance. These globally coordinated oscillations are not easily explained by ocean current mechanisms such as bistable flipping of ocean deep-water production or regional temperature changes involving the NW/SE migration of the North Atlantic polar front. They also are not accounted for by Earth orbital changes in seasonality or by increases in atmospheric $CO_2$ or $CH_4$. On the other hand, evidence of an elevated cosmic ray flux and of a major interstellar dust incursion around 15,800 years B.P. suggest that a cosmic ray wind driven incursion of interstellar dust and gas may have played a key role through its activation of the Sun and alteration of light transmission through the interplanetary medium.

## 1. Introduction

Climatic profiles from various parts of the world have been found to register synchronous climatic changes. Mörner (1973) has described evidence of correlated climatic fluctuations occurring during the past 35,000 years in Northern hemisphere climatic profiles and has concluded that they must have been global in extent. LaViolette (1983, 1987, 1990) later conducted an inter-hemispheric study which compared profiles from the British Isles (Atkinson et al., 1987), North Atlantic (Ruddiman et al., 1977), Gulf of Mexico (Leventer et al., 1983), and Southern Chile (Heusser and Streeter, 1980) and concluded that the Bölling-Alleröd/Younger Dryas (B/AL/YD) climatic oscillation occurred synchronously in both northern and southern hemispheres with the Bölling-Alleröd marking a period of global warming. Dansgaard, White, and Johnsen (1989) have compared oxygen isotope ($\delta^{18}O$) dated profiles from the Greenland Dye 3 ice core and a sediment core from Lake Gerzen, Switzerland and have shown that climatic oscillations during the B/AL/YD closely track one another in both cores. Also, Kudrass et al. (1992) have shown evidence of the B/AL/YD climatic oscillation in radiocarbon dated sediment cores from the Sulu Sea of Southeast Asia and note that it occurred contemporaneously with the B/AL/YD oscillation detected in a North Atlantic core. They note that the Younger Dryas cold period is recorded in radiocarbon dated cores from many parts of the world (e.g., Gulf of Mexico, North Pacific Ocean, Argentina, equatorial Atlantic, Bengal Fan) and conclude that it must be regarded as a global phenomenon.

To further evaluate the possibility that climate has varied in a globally synchronous manner over relatively short intervals of time, this paper compares dated climatic profiles from various parts of the world that span the Termination I boundary at the close of the last ice age. This boundary was chosen as the focus for this study because of the greater availability of well-dated, high-sample-density climatic profiles spanning this period. When considered together, these data indicate that climate at distant parts of the globe varied in a synchronous manner and imply that the Earth's thermal energy balance underwent major changes at the end of the ice age, and possibly on earlier occasions as well. Various mechanisms are examined to see whether any can account for such abrupt geographically coherent climatic changes.

## 2. Hemispheric Synchrony of the Terminal Pleistocene Climatic Oscillation

<u>Land and Sea Climatic Profiles</u>

Climate at the end of the last glaciation did not proceed irreversibly toward interglacial warmth, but rather was characterized by a sequence of interstadial-stadial oscillations (see Table I). The B/AL/YD climatic oscillation is apparent in radiocarbon dated climatic profiles from both Northern

Table I
Scandinavian Climatic Zone Dates

| Climatic Zone | Acronym | Calendar Date (Years B.P.) | C-14 Date (Years B.P.) |
|---|---|---|---|
| Preboreal warming | PB | 11,550 - 11,300 | 10,000 - 9,700 |
| Younger Dryas Stadial | YD | 12,700 - 11,550 | 11,000 - 10,000 |
| Alleröd Interstadial | AL | 13,800 - 12,700 | 12,100 - 11,000 |
| Older Dryas Stadial | OD | 13,870 - 13,800 | 12,150 - 12,100 |
| Bölling Interstadial | BO | 14,500 - 13,870 | 13,000 - 12,150 |
| Lista Stadial | LI | 14,850 - 14,500 | 13,300 - 13,000 |
| Pre-Bölling Interstadial | P-BÖ | 15,750 - 14,850 | 4,200 - 13,300 |

Calendar dates for these zones are based on dates assigned to corresponding climatic boundaries evident in the GRIP Greenland Summit ice core.

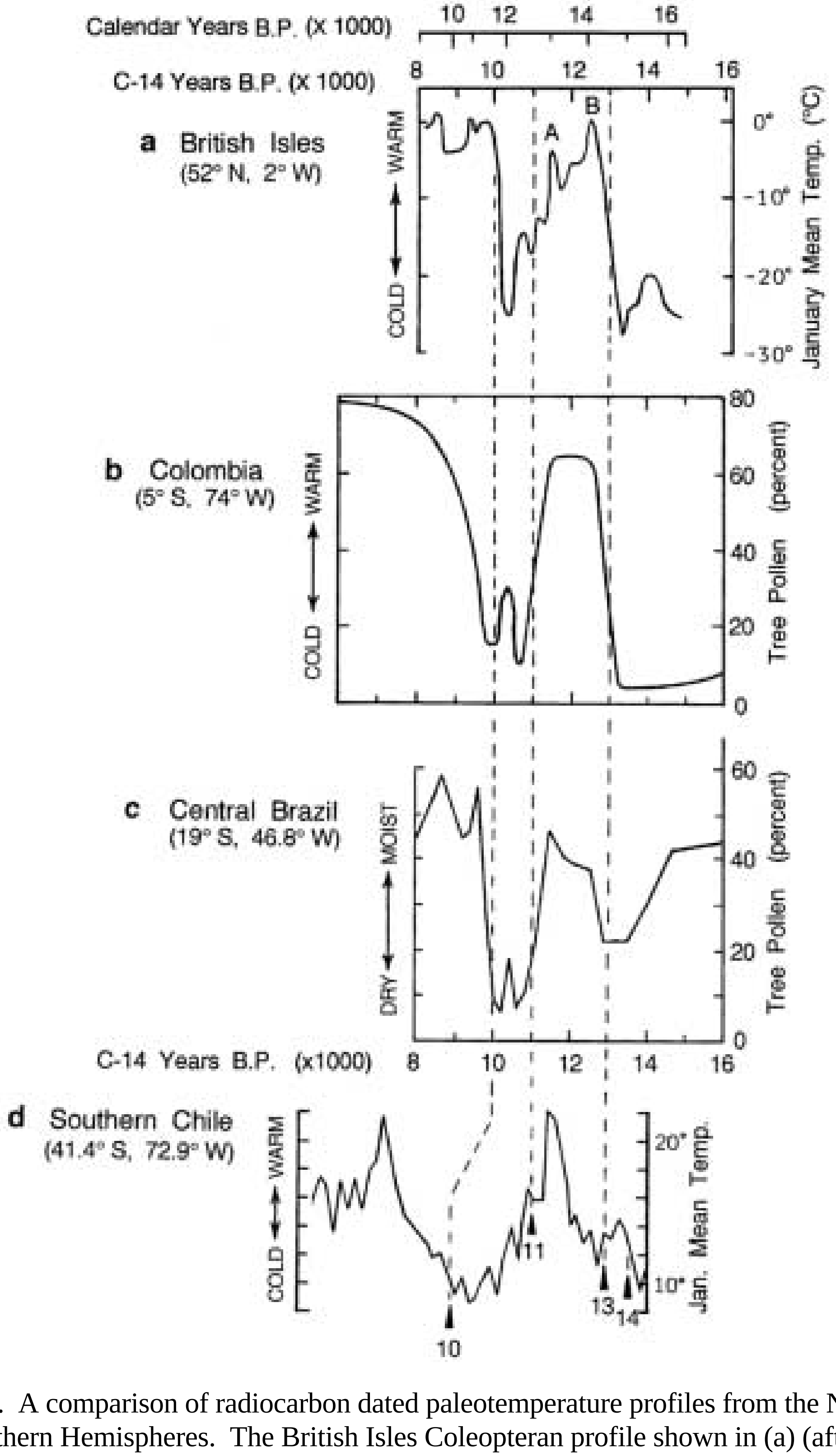

Figure 1. A comparison of radiocarbon dated paleotemperature profiles from the Northern and Southern Hemispheres. The British Isles Coleopteran profile shown in (a) (after Atkinson et al., 1987) is compared to pollen profiles from: (b) the El Abra Corridor, Colombia (after Schreve-Brinkman, 1978), c) central Brazil (after Ledru, 1993), and d) Alerce, Chile (after Heusser & Streeter, 1980).

and Southern Hemispheres (see Figure 1). The British Isles Coleopteran beetle profile (52° N, 2° W), shown in Figure 1-a, is time-calibrated with 49 radiocarbon dates (after Attkinson et al., 1987) and correlates well with the annual-layer-dated GISP 2 Greenland Summit ice core profile. The pollen diagrams from Colombia (5° S, 74° W), Central Brazil (19° S, 46.8° W), and Alerce, Chile (41.4° S, 72.9° W), Figures 1-b, -c, and -d (after Schreve-Brinkman, 1978; Ledru, 1993; Heusser and Streeter, 1980), are controlled by 20, 10, and 13 radiocarbon dates respectively. A comparison of these profiles indicates that this climatic oscillation was contemporaneous in these diverse locales. Climate in both hemispheres became unusually warm between 14.5 k to 12.7 k calendar years (cal yrs) B.P, equivalent to 13 k and 11 k $^{14}C$ yrs B.P.; see Table II for $^{14}C$ date conversions. During this period temperatures reached levels typical of the present interglacial, but cooled again to

Table II
Conversions from Radiocarbon to Calendar Dates

| Calendar Years B.P. | C-14 Years B.P | Correction Years |
|---|---|---|
| 11,050 | 9,500 | 1550 |
| 11,550 | 10,000 | 1550 |
| 12,100 | 10,500 | 1600 |
| 12,700 | 11,000 | 1700 |
| 13,300 | 11,500 | 1700 |
| 13,700 | 12,000 | 1700 |
| 14,200 | 12,500 | 1700 |
| 14,500 | 13,000 | 1500 |
| 15,100 | 13,500 | 1600 |
| 15,600 | 14,000 | 1600 |
| 16,000 | 14,500 | 1500 |
| 16,700 | 15,000 | 1700 |
| 17,300 | 15,500 | 1800 |
| 17,900 | 16,000 | 1900 |
| 19,000 | 17,000 | 2000 |
| 20,000 | 18,000 | 2000 |
| 21,000 | 19,000 | 2000 |
| 22,000 | 20,000 | 2000 |
| 27,000 | 25,000 | 2000 |
| 32,000 | 30,000 | 2000 |

The conversions of radiocarbon dates to calendar dates given in Table II were arrived at by correlating climatic horizons in radiocarbon dated land profiles with similar horizons evident in the GRIP and GISP2 ice core records which are dated with an absolute annual layer chronology (Johnsen et al., 1992; Taylor et al., 1993). The conversions for dates earlier than 15,000 $^{14}C$ yrs B.P., are based on a smoothed version of the $^{14}C$ dated uranium/thorium chronology of Bard et al.(1990a).

glacial levels during the Younger Dryas 12.7 k to 11.55 k cal yrs B.P. (11 k to 10 k $^{14}$C yrs B.P.). This cold period was then ended by the abrupt onset of the Preboreal warming which commenced the Holocene. Understandably, the ages for the climatic zone boundaries at a given site have some degree of error due to the uncertainty of up to several hundred years in any given radiocarbon date. Nevertheless, this uncertainty is small when compared with the relatively long duration of the B/AL warming (1950 cal. yrs) and YD cooling (1150 cal. yrs).

This same warming, cooling, and final rewarming is registered in ocean cores from different parts of the world. It is registered in the north in sediment core Troll 3.1 (60.8° N, 3.7° E) which plots foraminifera abundance in the Norwegian Sea as an indicator of sea-surface temperature; see Figure 2-a (after Lehman and Keigwin, 1992). It is also seen in Figure 2-b in a foraminifera profile from the Gulf of Mexico (21.0° N, 94.1° W) which charts the ratio of the warm water species *Globorotalia menardii* to the cold water species *Globorotalia Inflata* (after Beard, 1973). Again this oscillation is evident in Figure 2-c in the $^{14}$C dated foraminifera temperature profile SU 81-18 from the southeast coast of Portugal (37.8° N, 10.2° W) (after Bard et al., 1989) as well as in foraminifera $\delta^{18}$O profiles from cores penetrated in the India-Indochina equatorial region. These include a core from the Arabian Sea (15.5° N, 72.6° E), Figure 2-d (after Van Campo, 1986), a core from the Sulu Sea (8.2° N, 121.6° E), Figure 2-e (after Kudrass et al. 1991), and a core from the Bay of Bengal (11.8° N, 94.2° E), Figure 2-f (after Duplessy et al., 1981). A comparison of the radiocarbon dated profiles shown in Figures 2-a, -c, & -e indicates that, as in the land profiles, this Termination I boundary climatic oscillation was communicated to these widely separated regions with a minimal time lag.[1] During the Bölling-Alleröd sea-surface temperature off the Portuguese coast rose by 11°C to Holocene values (Figure 2-c). An increase to Holocene temperatures is apparent also in the Norwegian Sea and Gulf of Mexico profiles. The change in $\delta^{18}$O evident in the Sulu Sea core indicates that sea-surface temperatures changed by about 2 to 3°C, comparable to the glacial-interglacial temperature difference for this region (Kudrass et al., 1991).

In addition, the Younger Dryas cooling event has been detected outside of the Europe/North Atlantic region in a number of other studies: in the Gulf of Mexico (Flower and Kennett, 1990), in

---

[1] Ocean core $^{14}$C dates are typically revised by -440 years to bring them into conformance with land $^{14}$C dates, thereby correcting for the time lag involved in the entry of atmospheric $^{14}$C-laden $CO_2$ into the oceans. The standard correction was applied to $^{14}$C dates obtained for the profiles from Portugal and the Sulu Sea (Figure 2-b & -c). However, in the case of the Norwegian Sea core (Figure 2-a), a correction of -840 years must be applied in order to bring the $^{14}$C dates for its climatic horizons into conformance with dates for similar horizons observed in the British Isles Coleopteron profile located less than 1000 km away. The reason why radiocarbon dates at this northerly ocean location would require 400 years additional correction is unclear, but may be due to the influx of old atmospheric $CO_2$ from gases disolved in the incoming glacial meltwater and a lower rate of influx of young atmospheric $CO_2$ due to the presence of sea ice and a lid of low salinity water.

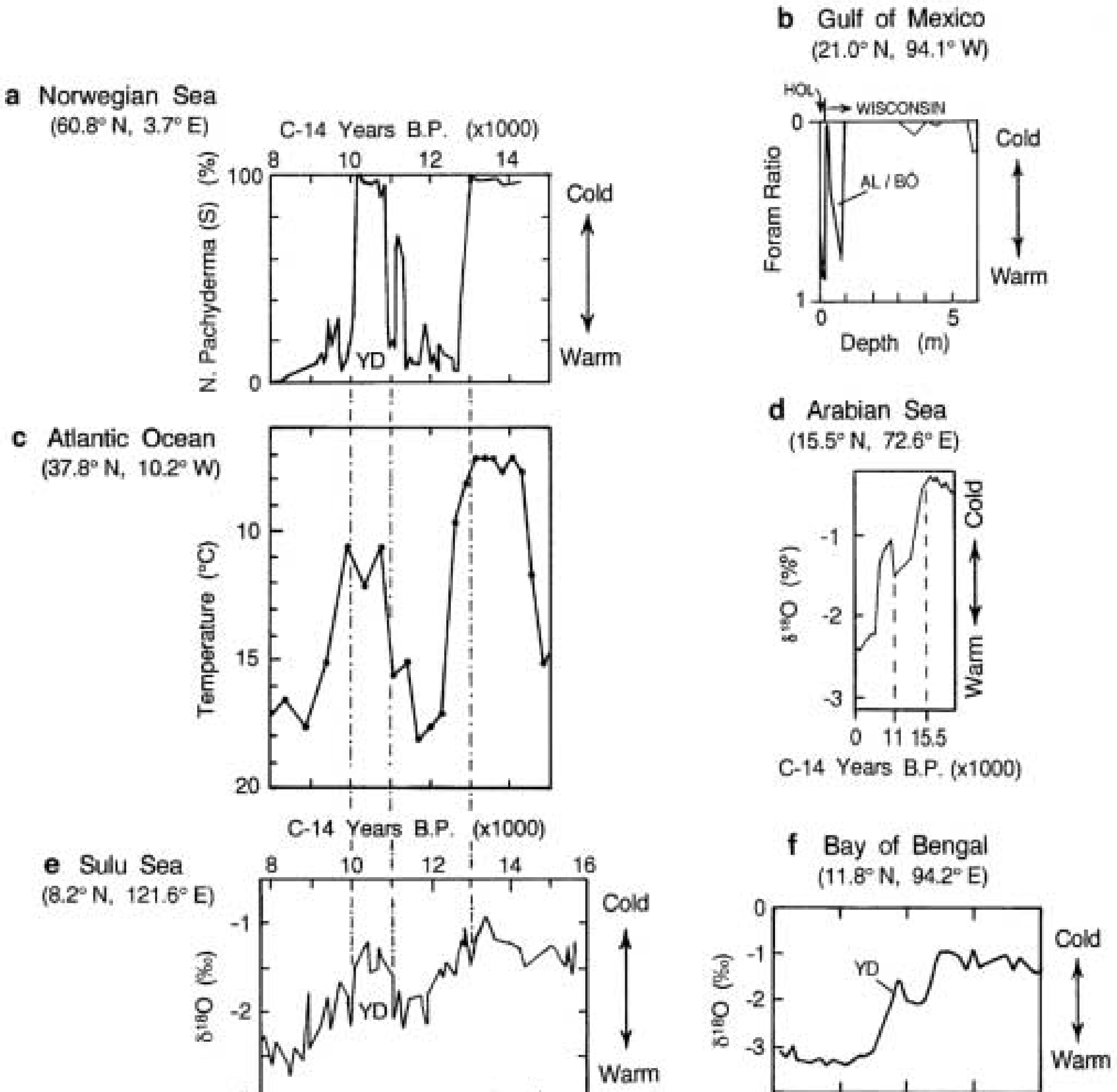

Figure 2. A comparison of ocean paleotemperature profile from various parts of the world: a) foraminifera abundance in Norwegian Sea core Troll 3.1 (Lehman and Keigwin, 1992), b) foraminifera ratio in Gulf of Mexico core 64-A-9-42 (Beard, 1973), c) foraminifera temperature profile SU 81-18 from the southeast coast of Portugal (Bard et al., 1989), d) $\delta^{18}O$ profile from Arabian Sea core MD 76-131 (Van Campo, 1986), e) $\delta^{18}O$ profile from Sulu Sea core SO49-82KL (Kudrass et al. 1991), and f) $\delta^{18}O$ profile from Bay of Bengal core MD 13-36 (Duplessy et al., 1981).

South America (Burrows, 1979; Harvey, 1980; Heusser and Rabassa, 1987; Heusser, 1984; Moore, 1981; Van der Hammen et al., 1981; Wright, 1984), Africa (Coetzee, 1967; Scott, 1982), in East Asia (Fuji, 1982), and in New Zealand (Burrows, 1979; Denton and Handy, 1994; Ivy-Ochs et al., 1999). Together, this evidence suggests that the Younger Dryas, and the Bölling-Alleröd interstadial that immediately preceded it, was of global extent.

The rapid onset and intensity of the Bölling-Alleröd global warming are not easily explained by terrestrial theories of climatic change. With the onset of the Bölling-Alleröd, winter temperatures in the British Isles increased by ~25°C and summer temperatures by 7 - 8°C to levels typically found in that local today (Figure 1-a). In southern Chile, summer temperatures warmed by 12°C, apparently reaching a level 7°C higher than the Holocene summer temperature mean. These warmings occurred at a time when the extensive continental ice sheet coverage kept the surface albedo of the glaciated regions about 50% higher than its present value (Budyko, 1974, pp. 279, 304). So, considering the relatively unfavorable solar energy balance conditions which then prevailed, a spontaneous amelioration of the Earth's climate comes as somewhat of a surprise. Whatever caused this global warming would have had to overcome this energy-balance handicap.

Evidence for the Bölling-Alleröd warming is also seen in the rapid melting of the ice sheets. The Scandinavian ice sheet began to recede rapidly northward at the onset of this interstadial, its recession rate reaching a maximum around 14,200 cal yrs B.P. and continuing at a somewhat lower rate through the Alleröd (Figure 3, lower profile). Ice sheet recession rate dropped dramatically with the onset of the Younger Dryas stadial, but surged upward again when this cold period was ended by the Preboreal warming.

Meltwater discharge from the North American ice sheet also reached a high level during the Bölling-Alleröd interstadial, as indicated by the high rate of freshwater discharge into the Gulf of Mexico (Kennett and Shackleton, 1975; Emiliani et al., 1978; Leventer et al., 1982, 1983). For example, the upper profile in Figure 3 plots $\delta^{18}O$ values for core EN32-PC4 penetrated in the northwestern Gulf of Mexico Orca Basin (26.9°N, 91.4°W) (Broecker et al., 1989). The shaded region, characterized by excessively negative $\delta^{18}O$ values, indicates a time when the Mississippi River was rapidly discharging isotopically light glacial meltwater into the Gulf. The magnitude of the isotopic excess reflects the rate of meltwater discharge, which in turn depends upon the temperature environment in the vicinity of the North American ice sheet and the fluvial routing of the meltwater.

The Orca profile indicates that meltwater discharge into the Gulf ceased during the Younger Dryas, but resumed once again with the onset of the Preboreal warming. Ice sheet recession rate in Scandinavia underwent a similar decrease and resurgence about this same time. This correlated behavior suggests that glacial melting responded in a similar way on both sides of the Atlantic and as a response to the prevailing change in air and ocean temperature which was warm during the Bölling-Alleröd, cold during the Younger Dryas, and warm again during the Preboreal. A decrease in North American ice sheet meltwater output during the period 12,700 to 11,550 cal yrs B.P. (11 k - 10 k $^{14}C$ yrs B.P.) is consistent with the global onset of the Younger Dryas cold interval. Evidence that this cold period occurred in the Gulf regions is seen in Figure 2-e (Beard, 1973) and in the more accurately dated core EN32-PC4 (Flower and Kennett, 1990). It also is consistent with similar changes in temperature and ice accumulation rate evident in Greenland ice cores (Dansgaard et al., 1982; Johnsen et al., 1992, Taylor et al., 1993, Alley et al., 1993).

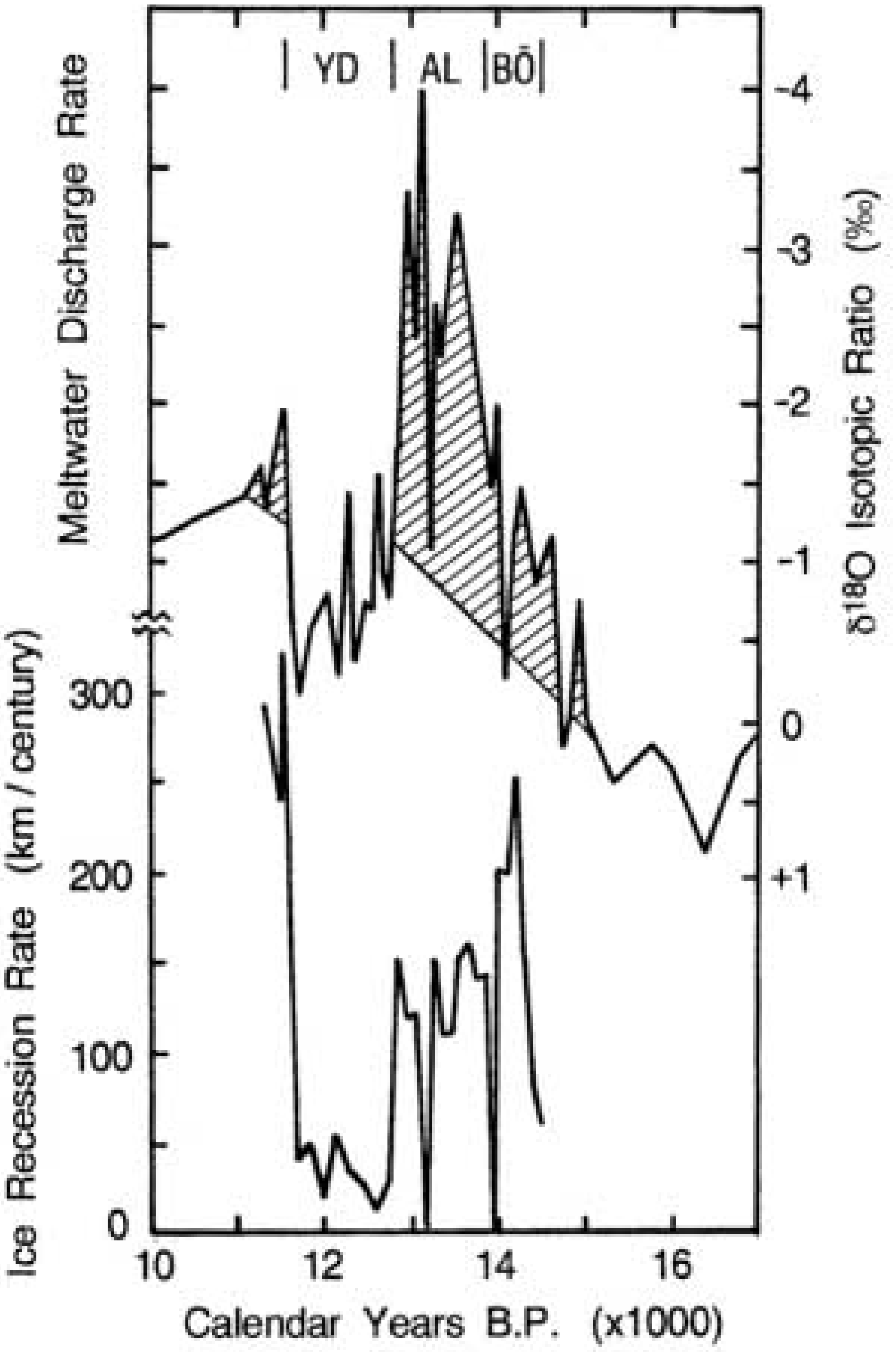

Figure 3. Upper profile: $\delta^{18}O$ profile for Gulf of Mexico Orca Basin core EN32-PC4 (after Broecker et al., 1989). Shaded portion charts the rate of meltwater discharge into the Gulf of Mexico. Lower profile: ice sheet recession rate in southern Sweden based on data taken from Björck and Möller (1987) and Tauber (1970). Climatic zones: Younger Dryas (YD), Alleröd (AL), and Bölling (BO).

One theory attributes the cessation of meltwater input into the Gulf during the Younger Dryas primarily to a diversion of the meltwater routing away from the Mississippi River and into the St. Lawrence as the retreating ice sheet removed the glacial blockage of the eastern outlet of Lake Agassiz (Broecker et al., 1989). This theory further suggests that discharge down the Mississippi again recommenced for a short period during the Preboreal when this eastern outlet was again blocked by the Marquette glacial advance. However, the finding that discharge into the Gulf ceased at a time when regional climate abruptly cooled and glacial melting halted and then recommenced at a time when regional climate abruptly warmed up again and glacial melting had begun again suggests an obvious cause-effect relation. While meltwater diversion must have played some role during this period, for the most part the Gulf record appears to be charting the melting rate response of the North American ice sheet to global changes in climate.

Times of maximal rate of sea level rise should be expected to correlate with periods of global warming. In fact, as seen in the Barbados coral reef record (Figure 4), the rate of sea level rise peaks during both the Bölling-Alleröd (meltwater pulse IA) and during the Preboreal (meltwater pulse IB). This suggests the ice sheets were collectively discharging meltwater at a maximal rate during these times and hence that these climatic ameliorations were not geographically localized. The Barbados record also indicates that meltwater discharge rate was low during the Younger Dryas, thereby supporting the point made earlier that the Gulf of Mexico cessation event was largely due to a global cooling and not to a regional redirection of the Laurentide meltwater outflow to the St. Lawrence River.

Times when sea level was rising at a maximum rate (Figure 4, IA and IB) match up quite well with times of high ice sheet recession rate evident in Scandinavia during the Bölling and Preboreal warm phases (Figure 3, upper profile). Peak IA of the global meltwater discharge rate profile, which began its rise at around 14,400 cal yrs B.P. and peaked around 14,000 cal yrs B.P., lagged by about 200 years compared to peak IA of the Scandinavian ice recession rate record. This lag suggests that the early stage of deglaciation was dominated by melting of the marine-based parts of the ice sheet, which contributed little to sea-level rise (see Veum et al., 1992). A closer correlation is apparent with Meltwater pulse IB which began its rise at the beginning of the Preboreal around 11,600 cal yrs B.P. and declined around 11,000 cal yrs B.P.

Climate profiles from the British Isles and southern Chile both record a minor warming event, prior to the Bölling, spanning the period 15,750 to 14,850 cal yrs B.P. (14.2 to 13.3 k $^{14}C$ yrs

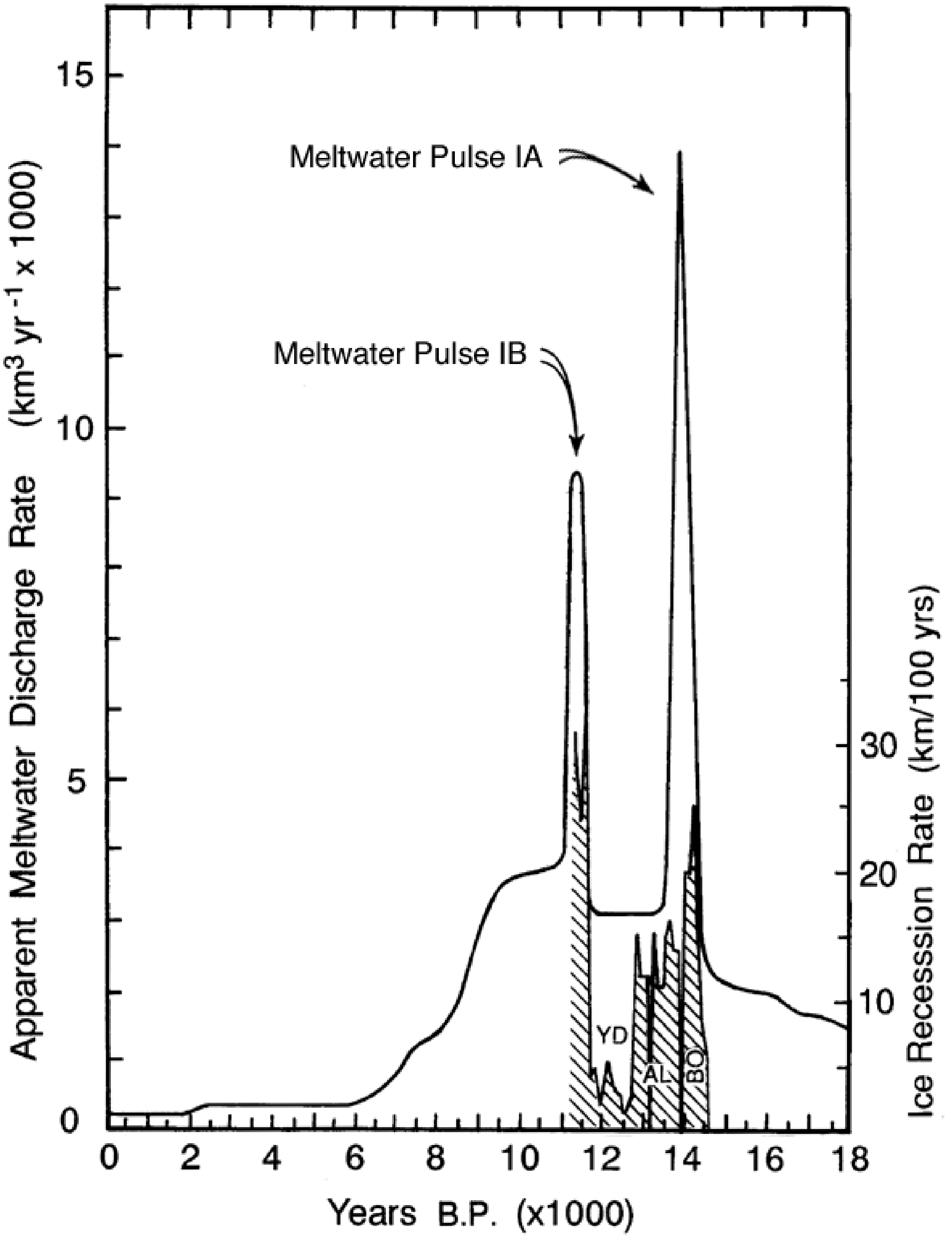

Figure 4. Upper profile: the rate of global glacial meltwater discharge into the oceans calculated from the Barbados sea level change curve (Fairbanks, 1989), revised according to the U-Th $^{14}C$ calibrations of Bard et al. (1990a, 1990b). Lower profile: ice sheet recession rate in southern Sweden. Climatic zones: Younger Dryas (YD), Alleröd (AL), and Bölling (BO).

B.P.); compare Figures 1-a and 1-d. This correlates in the East Baltic area with the Msta and Raunis interstadials (ending ~13.25 k $^{14}$C yrs B.P.), with the Susaca interstadial of Columbia (Dreimanis, 1966), and in the Great Lakes Region with the Mackinaw (or Cary-Port Huron) interstadial (13.3 ± 0.4 k $^{14}$C yrs B.P.) and with the earlier warm period that preceded the deposition of the Wentworth till. Although given different names in different regions, this "Pre-Bölling" interstadial appears to have been of global scope, although not nearly as intense as the Bölling-Alleröd.

The several hundred year long cool interval that separated this warm period from the Bölling, evident in the British Isles and Chilean records, correlates with the Lista and Holland Coast stadials in Southern Norway and Sweden (13.5 - 13.0 k $^{14}$C yrs B.P.) (Berglund, 1979) and with the Luga stadial in the Baltic area (13.2 - 13.0 k $^{14}$C yrs B.P.) (Raukas and Serebryanny, 1972; Berglund, 1979). In the Great lakes area, this cooling matches up with the Port Huron stadial, which dates at 13,000 ± 500 $^{14}$C yrs B.P. and divides the Mackinaw from the Two Creekan interstadial (Karrow, 1984; Dreimanis and Goldthwait, 1973). Thus climatic oscillations occurring between 15,750 and 14,500 years ago (14.2 - 13.0 k $^{14}$C yrs B.P.) also show evidence of transatlantic and interhemispheric correlation.

Ice Core Climatic Profiles

The Earth's polar ice record also contains evidence of globally correlated climatic changes. The B/AL/YD oscillation, for example, is synchronously registered in both the GISP2 Summit, Greenland and Taylor Dome, Antarctica ice core climate profiles; see upper two profiles in Figure 5. Steig et al. (1998) measured atmospheric methane concentration from air bubbles trapped in the ice and used the observed rapid concentration changes as markers for correlating the two deuterium isotope climatic records. This matching showed that the climatic transition boundaries correlate closely in time amongst the two cores. Climate at the Taylor Dome site began to gradually warm around 15,500 years BP and experience a more rapid warming around 14,600 years B.P., synchronous with the beginning of the Bölling warming registered in the Summit, Greenland ice core profile. A cold spike at around 13 kyrs BP at Taylor Dome is correlative with the Intra-Alleröd Cold Peak at Summit, Greenland. The subsequent Younger Dryas cool period is not as distinctive at Taylor Dome as it is at Summit. However, the sudden warming at around 11.7 kyrs BP registered in the Taylor Dome core is correlative with the abrupt warming registered at Summit at the beginning of the Holocene.

Climatic synchrony is also evident between the Summit, Greenland and Byrd Station, Antarctica ice core isotope records; see Figure 5. The isotope profile for the Byrd core (Johnsen et al., 1972) is dated according to the chronology of Beer et al. (1992) which they obtained by correlating distinctive $^{10}$Be concentration peaks found in both the Byrd Station, Antarctic and Camp Century, Greenland isotope records, some peaks dating as early as 12 - 20 kyrs BP. The Camp Century isotope profile, in turn, has been accurately dated through correlation with the annual layer dated Summit, Greenland isotope profile (Johnsen et al., 1992). The Byrd oxygen isotope profile shows a progressive climatic amelioration beginning around 15,800 calendar years BP, correlative with the beginning of the Pre-Bölling Interstadial evident in the Summit, Greenland core, and continuing through the Bölling and Alleröd Interstadials. The cooling evident from 1143 to 1100 meters core depth, termed the Antarctic Cold Reversal (ACR), dates between 13,250 to 11,600 calendar years BP and correlates with the Intra-Alleröd Cold Peak and Younger Dryas registered in the Summit record. The Byrd record shows this cooling more clearly than the Taylor Dome record. A terminal warming is evident in the Byrd core around 11.6 kyrs BP correlative with the beginning of the Holocene PreBoreal in the Northern Hemisphere. The $^{10}$Be correlations of Beer et al. contradict the conclusions of Blunier, et al. (1998) that the ACR at Byrd Station had occurred 500 years prior to the Younger Dryas.

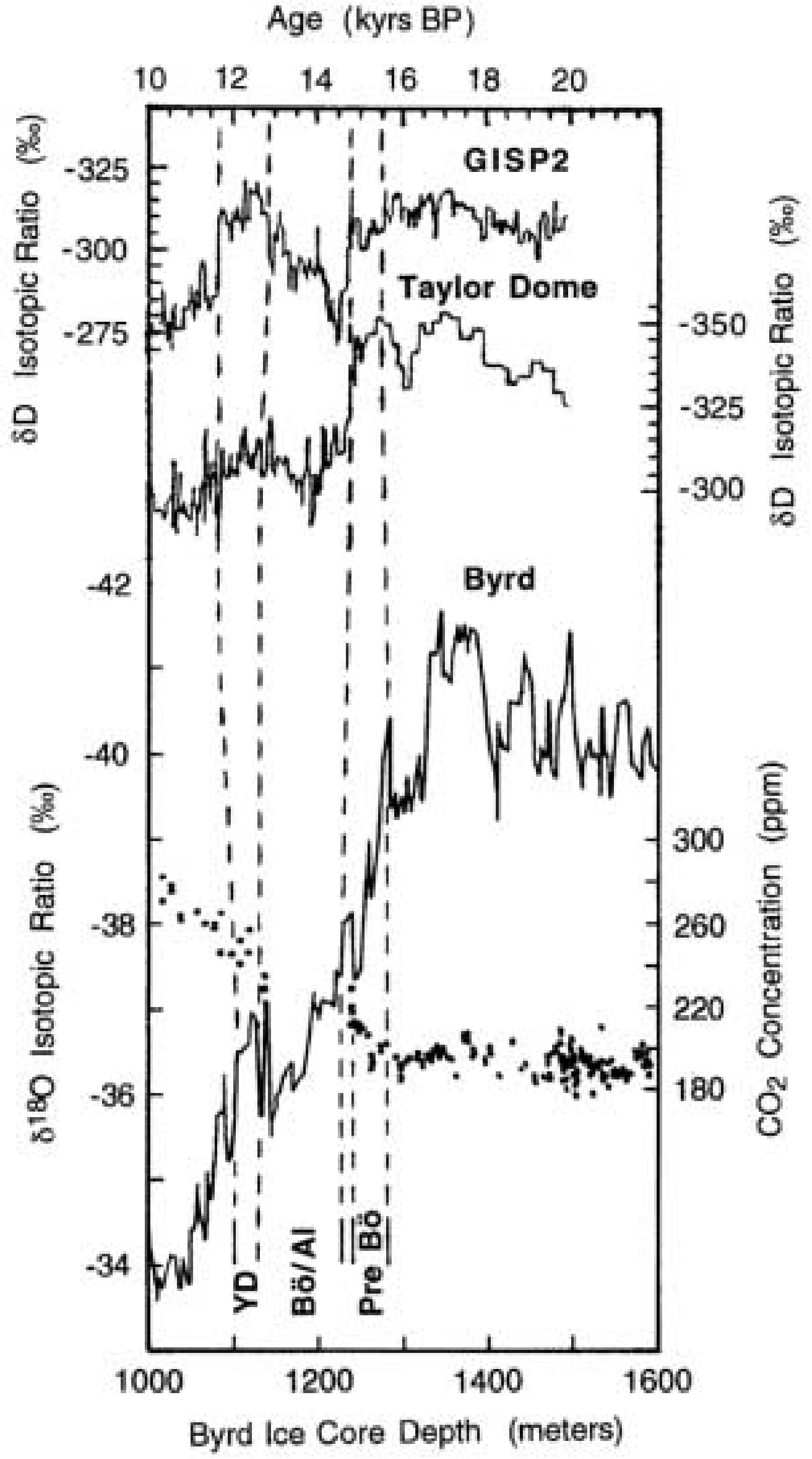

Figure 5. A comparison of Greenland and Antarctic ice core profiles showing climatic synchrony of the Bölling-Alleröd-Younger Dryas oscillation. Upper profiles: Summit, Greenland GISP2 deuterium profile correlated to the Taylor Dome, Antarctic deuterium profile using methane as an indicator (adapted from Steig et al., 1998). Lower profile: Byrd Station, Antarctica $d^{18}O$ profile (Johnsen et al., 1972) correlated to the Greenland ice record by means of Be-10 peaks (Beer et al., 1992). The $CO_2$ data is taken from Neftel et al. (1988).

Jouzel et al. (1992) have developed a chronology for the Vostok and Dome C Antarctic ice cores by using a two-dimensional flow model along with a saturation vapor pressure approach that they base on ice core isotope data. Their technique leads to less than a 3% error in dating the 35 kyrs BP Be-10 peak registered in each ice core. Although these chronologies predict differing dates for the beginning of the ACR (11.9 kyrs BP at Vostok and 13.4 kyrs BP at Dome C), they conclude that the apparent 1500 year phase lag they had originally calculated for the Vostok ACR is not real, but due to dating inaccuracy, and they note that the assumption of climatic synchrony has the advantage that the beginning of the "Holocene" dust concentration minimum in each core is made contemporaneous. Jouzel et al. adopt the Dome C date of 13.4 kyrs BP as the correct date for the beginning of the ACR., which corresponds closely with 13.2 kyrs BP date for the beginning of the ACR at Byrd Station and for the beginning of the Intra-Allerod cold peak at Summit, Greenland. This synchrony is corroborated by the findings of Mulvaney et al. (2000). Using Ca concentration to synchronize the Taylor Dome and Dome C ice core isotope profiles, they argue that at least the ACR feature occurred synchronously at both locations, hence that climate in various parts of Antarctica changed in a synchronous manner.

<u>Problems with the Argument for Asynchronous Climatic Change</u>

Others have concluded that climatic changes were asynchronous among different parts of Antarctica and also between Antarctica and Greenland. For example, Blunier et al. (1998) have derived a different time scale for the Byrd and Vostok ice cores by pegging their profiles to the annual-layer-dated Summit profile using excursions of atmospheric methane as inter-core markers. From this they conclude that the ACR began around 13.8 kyrs BP at Byrd Station, 600 years prior to the Intra-Alleröd Cold Peak in Summit, Greenland, but that the ACR at Vostok began around 15.0 kyrs BP, 1200 years prior to the Byrd Station ACR. Thus Blunier et al. propose that climate at Vostok cooled and then warmed up again, that 1200 years later climate at Byrd Station (about 4000 km away) similarly cooled and then warmed up, and that 600 years later climate at Taylor Dome (about 2700 km from Byrd Station) similarly cooled and then warmed up again. This would require some sort of exotic refrigeration mechanism proceeding at Vostok while the ice age was in the process of ending at Byrd Station, and that later was capable of cooling Byrd Station while the ice age was in the process of ending at Taylor Dome. Even greater age discrepancies are projected for the Bölling deglacial warming, being professed to begin in Vostok and Byrd Station around 17,000 to 18,000 years BP and to begin at Taylor Dome about 3000 years later at 14,500 years BP.

In reporting their finding that the Taylor Dome Antarctic core registers the AL/BO/YD oscillation in synchrony with the north, Steig et al. (1998) also adopt the chronology of Blunier et al. (1998) with its implication that the deglacial warming at the Byrd and Vostok sites was asynchronous and that it preceded the climatic warming in Greenland. Steig et al. concluded that the 2000 year time lag between the 13.0 kyrs BP Antarctic Cold Reversal at Taylor Dome and the date projected for the Vostok ACR is real. They suggested that the Byrd and Vostok sites failed to synchronize with the Northern Hemisphere climatic phases because these sites lay further from open water. However, to the contrary, during the last ice age the Taylor Dome, Byrd, and Vostok sites were all approximately equidistant from the outer sea ice boundary.

Moreover since these three sites lie within 3000 kilometers of one another, it does not make sense that they were climatically isolated from one another and registered asynchronous changes. Just as the isotope profiles of the Dye 3, Summit, and Camp Century, Greenland sites, which lie within 1500 km of each other, have been shown to register synchronous climatic changes (Johnsen et al., 1992), so too these various Antarctic sites should have been exposed to similar climatic conditions. But the chronology of Blunier et al. implies that the Byrd and Vostok sites had been experiencing near interglacial warmth for more than 1000 years while the Taylor Dome site had been maintaining full glacial conditions. Instead, it is far more likely that these phase lags are artifacts arising from inaccuracies in the ice core chronologies.

For example, the technique of using methane concentration for correlating ice cores has the inherent uncertainty that the difference in age between the sampled air bubbles and their surrounding ice matrix ($\Delta_{age}$) is not a known measured quantity. The magnitude of this difference depends on the estimated rate of ice accumulation and on the estimated depth at which air became sealed off into bubbles when the firn compacted to form ice. The estimate of this seal-off depth can vary depending on a number of factors. The calculations, which must be done separately for each ice core, are model dependent and highly assumption laden.

In view of the land and ocean core evidence reviewed earlier which indicates synchronous globally climatic change at the end of the last ice age, we are inclined to adopt the chronologies of Beer et al. (1992) and Jouzel et al. (1992) over that of Blunier et al. Combined with the findings of Steig et al. (1998) on the synchrony of deglaciation in Summit, Greenland and Taylor Dome, Antarctica, these various chronologies lead to the conclusion that the B/AL/YD climatic oscillation recorded in Greenland ice occurred synchronously with similar climatic changes registered in various parts of Antarctica and tracked climatic changes occurring in other northern and southern hemispheric regions. With the conclusion of climatic synchrony, the dating which Blunier propose for the Vostok, Dome C, and Byrd Station for the period 14.5 -19.5 kyrs BP would be made younger and would require that precipitation over this deglacial warming period was higher than they had supposed. However, this is entirely expected since ice accumulation rate is known to be high at times of warming.

## 3. The Apparent Inadequacy of Terrestrial Explanations

Amplified Fluctuations

The ice accumulation rate profile from the GISP2 Summit core indicates that the climatic warming from the Younger Dryas to the Preboreal occurred within a few years time and that the warming from the Older Dryas to the Bölling occurred almost as rapidly (Alley et al. 1993). At present there is no general consensus as to the cause of such abrupt climatic changes. Milankovitch precessional and nutational cycles have periods of the order of 20 to 40 thousand years and hence, by themselves, cannot account for the rapidity of the terminal Pleistocene climatic oscillations. It has been suggested that slowly varying changes in seasonality might bring the climatic system past a certain critical point where nonlinear positive-feedback processes encourage random fluctuations (e.g., weather noise) to rapidly grow in size and drive ice sheet area and global climate to a new stable equilibrium (North and Crowley, 1985). However, since the climatic system incorporates negative feedback relationships which give it some degree of stability and tend to maintain it in a given climatic state, be it glacial or interglacial, destabilizing perturbations must exceed a certain critical size if they are to effect any large-scale change. Those that are too small in magnitude or duration will fail to change the system's prevailing climatic state. Weather noise probably belongs to this subcritical category.

Another point to consider is the global nature of the Bölling- Alleröd warming. Theories proposing that this was seeded from indigenous climatic fluctuation arising in a specific locale (e.g., in the North Atlantic) presuppose that it subsequently was rapidly communicated to other parts of the globe. For this to occur, positive-feedback processes would have to amplify the original perturbation sufficiently fast that the entropy-increasing tendency of geographic dispersal would be counteracted. But it is not clear what positive-feedback process could have operated on indigenous thermal fluctuations to warm climate around the world to the extent of increasing the rate of glacial melting by six fold within a matter of just a few hundred years, as occurred during the Bölling. Moreover it is also not clear why this process would suddenly shut off and allow global climate to temporarily relapse back to a glacial mode, as had occurred with the onset of the Younger Dryas. Instead, the circumstances call for a geographically diffuse mechanism capable of simultaneously affecting the energy balance of the entire planet.

$CO_2$ Greenhouse Warming

Seasonality changes produced by gradual Milankovitch orbital cycle variations may affect the Northern Hemisphere to some extent, but have little effect on the Southern Hemisphere. Hence they are unable to account for the hemispheric synchronism of glacial terminations (Manabe and Broccoli, 1985).

It has been suggested that global synchrony might have been achieved through some kind of interhemispheric linking, such as changes in atmospheric $CO_2$ concentration (Corlis, 1982; Manabe and Broccoli, 1985; Johnson and Andrews, 1986). However, by itself, $CO_2$ produces a relatively small greenhouse warming effect. For example, the Vostok ice core measurements of Barnola et al. (1987) show that at the end of the ice age $CO_2$ concentration rose by 25% from 195 ppm to 260 ppm. The increased IR opacity resulting from this rise would have produced a warming of only 0.4° C, contributing only 5% of the total 9° C temperature increase (Genthon et al., 1987). Moreover the Byrd ice core data indicates that $CO_2$ concentration continued to increase through the Younger Dryas, just the opposite of what would be expected if $CO_2$ played a critical role in modulating climate (see Figure 5). In summary, there is little evidence to suggest that the rise in atmospheric $CO_2$ concentration was the cause of the Termination I global warmings. Rather, the rise in atmospheric $CO_2$ was more likely a response to global warming, as the warming oceans released their dissolved gas to the atmosphere.

Compared with carbon dioxide, methane underwent a much larger percentage increase at the end of the ice age, doubling from about 360 ppb to 725 ppb, as determined from measurements of the Summit, Greenland ice core (Chappellaz et al, 1993). However, since its absolute concentration is 1000 fold less than that of $CO_2$, it is not a major contributor to greenhouse warming. Rather, its increase also is most likely a response to climatic change rather than an instigator, the rise in $CH_4$ concentration being attributed to the increased abundance of vegetation which is a major producer of this gas.

Deep-Ocean Circulation

Broecker et al. (1985, 1988a, 1989, 1990) have proposed that the abrupt warming registered in the North Atlantic around 14,650 cal yrs B.P. (13.0 k $^{14}C$ yrs B.P.) was produced by a change in the rate of North Atlantic deep-water (NADW) production. This theory suggests that during the last ice age cool surface waters prevailing in the North Atlantic reduced the rate of evaporative loss there and thereby lowered the production rate of salty deep-water. This, in turn, would have caused the ocean-current "conveyor belt," which transports cold salty deep waters to the North Pacific and warm equatorial surface waters to the North Atlantic, to operate at a very minimal level. This would have cut off the supply of ocean heat feeding the Northern Atlantic atmosphere and, in so doing, would have helped to stabilize the prevailing glacial conditions. The theory goes on to suggest that the maximum seasonality (hot summers and cold winters) prevailing in northern latitudes toward the end of the last ice age increased evaporative loss and deep water production sufficiently to cause NADW production to rapidly flip to its high-flux interglacial mode. The warm equatorial water said to have been brought into the North Atlantic is theorized to have ameliorated climate in this region sufficiently to have induced the Bölling temperature rise. Furthermore, with the opening of the St. Lawrence River drainage system, an increasing influx of low-salinity meltwater is theorized to have temporarily returned NADW production to its glacial mode and thereupon induced the Younger Dryas cooling.

However, studies of benthic foraminifera in the Atlantic suggest that NADW production did not flip to its interglacial high-flux mode until around 12,500 $^{14}C$ yrs B.P., or about 500 $^{14}C$ years after the onset of the Bölling (Jansen and Veum, 1990; Veum et al., 1992; Charles and Fairbanks, 1992). So, the onset of NADW production cannot be the agent that caused the rapid warming at the beginning of the Bölling.

Moreover, the finding that tropical surface waters warmed during the Bölling-Alleröd interstadial is problematic for the deep-water circulation theory given that the proposed renewal of NADW circulation would have removed a substantial amount of heat from the equatorial region. For example, it is estimated that ocean currents presently transport northward about 1.4 X $10^{15}$ watts of heat annually, which amounts to about 1% of the annual solar irradiance (Stommel, 1980; Berger, 1990). All other things being equal, this heat removal should have decreased the temperature of equatorial surface waters, but, instead, an increase is seen (Figure 2). To adequately explain the Bölling-Alleröd and Preboreal global warmings what is needed is a mechanism that can rapidly increase the heat budget of the entire planet, as opposed to just redistributing the existing heat.

Moreover changes in NADW production also fail to explain the occurrence of the Younger Dryas. The Barbados sea-level profile indicates that the global rate of meltwater discharge was reduced during the Younger Dryas (Figure 4). Presumably, the meltwater flow into the North Atlantic also was lower during this time despite the possible opening of the St. Lawrence discharge route (Fairbanks, 1989). Thus NADW production is less likely to have shut down during that period. In fact, benthic evidence from the Norwegian Sea suggests that modern-type ocean circulation operated during the Younger Dryas (Veum et al., 1992). Also, $^{13}C$ data from the North Atlantic indicates that NADW production fluctuated greatly during both the Bölling-Alleröd and Younger Dryas deglaciation phases with no evidence of an additional prolonged slowdown occurring during the Younger Dryas (Berger, 1990). So the evidence rules against seeking a cause for the Younger Dryas in a thermohaline convection mechanism.

Moreover the carbon isotope record indicates a shut down of deep-water production during the Preboreal warming, a time when the rate of meltwater discharge to the oceans had reached a maximum (Berger, 1990; Boyle and Keigwin, 1987). This is consistent with the theory that large influxes of low-salinity meltwater reduce NADW production. However, since less heat would have been advected to the North Atlantic, this shut down should have opposed the Preboreal warming, rather than promoted it. While NADW production attractors could very well play a role in stabilizing the Earth's climate in the glacial or interglacial mode, ocean circulation changes are unable to account for the abrupt onset of Termination I warmings and coolings which occurred in coordinated fashion in diverse parts of the globe.

Polar Front Migration

The climatic evidence presented earlier also does not support theories that attribute the Termination-I warming to geographically localized effects such as the NW-SE migration of the North Atlantic polar front discussed by various authors (Ruddiman et al., 1977; Ruddiman and McIntyre, 1981; Mercer and Palacios, 1977). Such a regional mechanism would not account for the seemingly correlated oscillations in temperature and glacial wastage that occurred in various locations around the world, including the Indian/Indochinese tropics. Moreover the theory encounters difficulties in the North Atlantic as well. As Atkinson et al. (1987) point out, Great Britain's climate began to cool as early as 12,200 ± 200 $^{14}$C yrs B.P., long before the cold waters of the polar front began to return to their southerly position.

## 4. A Possible Galactic Explanation

Galactic Cosmic Ray Volleys.

The dramatic climatic shifts that took place during the Pleistocene may have had an extraterrestrial cause. One indication comes from the occurrence in ice age polar ice of high concentrations of $^{10}$Be, a 1.5 Myr half-life isotope generated when cosmic ray protons impact nitrogen and oxygen nuclei in the atmosphere (Raisbeck et al., 1981, 1987; Beer et al., 1984a, 1985, 1988, 1992). When adjusted for changes in ice accumulation rate, $^{10}$Be profiles can provide useful information that can help us determine if cosmic ray intensity has varied in the past. For example, the profiles shown in Figures 6 and 7 suggest that the cosmic ray background intensity was quite high on several past occasions. As explained in Appendix A, these profiles were calculated by multiplying the $^{10}$Be concentration found in polar ice at a particular depth by the corresponding ice accumulation rate to determine the atmospheric $^{10}$Be production rates which is directly correlated with cosmic ray intensity striking the Earth's atmosphere. The values were then normalized relative to Holocene values to produce relative cosmic ray intensity profiles. To get the unmodulated cosmic ray intensities, an additional adjustment must be made for solar magnetic screening which is dependent on the level of solar activity.

In an attempt to more conservatively explain these peaks as arising solely from terrestrial causes, some have suggested that they have been produced by variations of the geomagnetic field, suggesting that $^{10}Be$ production is higher when the geomagnetic field is at a minimum allowing an increased background cosmic ray flux to penetrate to the atmosphere. However, Beer et al. (1984b, 1988) find that the geomagnetic field has little effect on $^{10}Be$ variations. They report that $^{10}Be$ concentration in Camp Century, Greenland ice remained relatively constant between 0 and 4000 BC, despite a 40 percent decrease in geomagnetic dipole intensity. This is not surprising since the high energy cosmic rays responsible for $^{10}Be$ production are not easily screened by the Earth's magnetic field, especially in the polar regions. For example at 0° latitude about 20% of the energy flux of cosmic rays in the 3 to 10 Gev energy range would be screened. At 30° latitude, screening would drop to 10%, and at 77° latitude, where Camp Century is located, screening would be negligible. So if $^{10}Be$ deposited in polar ice originates in the local atmosphere, $^{10}Be$ variations found in polar ice records should be immune from variations in geomagnetic field intensity.

Raisbeck et al. (1987) have suggested that some peaks in the $^{10}Be$ record could be local enhancements that resulted from changes in atmospheric flow patterns which may have locally concentrated the isotope. However, their proposal is countered by the work of Beer et al. (1992) who have found that at least two major peaks in the $^{10}Be$ record appear both in both the Greenland and Antarctic ice records and therefore reflect actual enhancements in the rate of atmospheric $^{10}Be$ production. For example, they find that the 35,000 year old $^{10}Be$ peak in the Vostok Antarctic ice core (at (600 m) also appears in the Dome C and Byrd Station Antarctic records (at 830 m and 1750 m) and in addition appears in the Camp Century, Greenland ice core (at ~1218 m log book depth). Also they have located a 23,000 year old $^{10}Be$ peak in the Byrd core (at ~1500 m) which correlates with a similar peak located in the Camp Century core (at ~1190 m).

Interstellar cosmic rays more easily penetrate the heliopause magnetic sheath during times in the sunspot cycle when solar flare activity is at a minimum and thereby expose the Earth to elevated cosmic ray intensities. For example, during recent solar flare minima $^{10}Be$ production increased by up to 60% above its mean level (Beer et al., 1985). However, solar modulation of this magnitude does not account for Pleistocene $^{10}Be$ peaks that often rise several times higher than this. For example, the cosmic ray intensity profile presented in Figure 6, which adjusts $^{10}Be$ concentration for changes in ice accumulation rate, shows about a dozen peaks that display an increase of over 100% above the Holocene background level. Moreover explaining some of the less prominent $^{10}Be$ peaks in terms of reduced solar modulation would require periods of solar flare dormancy lasting several thousand years, over an order of magnitude longer than the Maunder Minimum. Although it could be argued that the Sun endured such long periods of inactivity during the ice age, several sets of data instead suggest that solar flare activity at the end of the ice age was instead much higher than it is at present (see next subsection). If so, the magnitude of the Bölling-Alleröd $^{10}Be$ production rate peak may be underestimated as a result of excessive solar modulation. So through a process of elimination, it may be concluded that the higher $^{10}Be$ peaks evident in the polar ice record register times when the background cosmic ray flux was particularly enhanced.

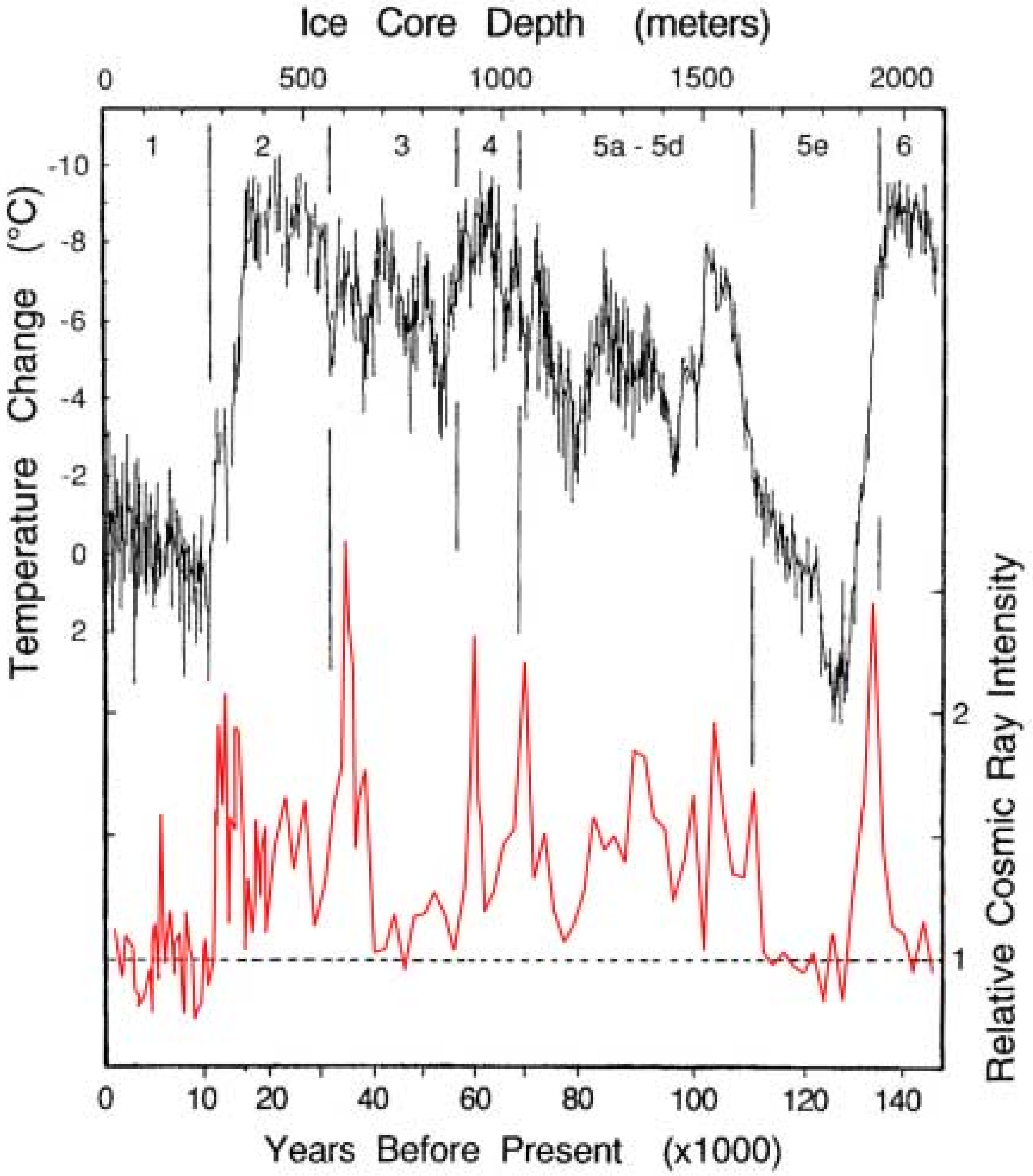

Figure 6. Lower profile: Cosmic ray intensity impacting the solar system (0 – 145 kyrs B.P.) normalized to present levels (based on the Vostok, Antarctica ice core $^{10}Be$ concentration data of Raisbeck et al. [*The Last Deglaciation*, p. 130], adjusted for changes in ice accumulation rate and solar wind screening and normalized to the Holocene average; see Appendix A, Part A). Upper profile: Ambient air temperature, as indicated by the ice core's deuterium content (from Jouzel, *Nature*, p. 403).

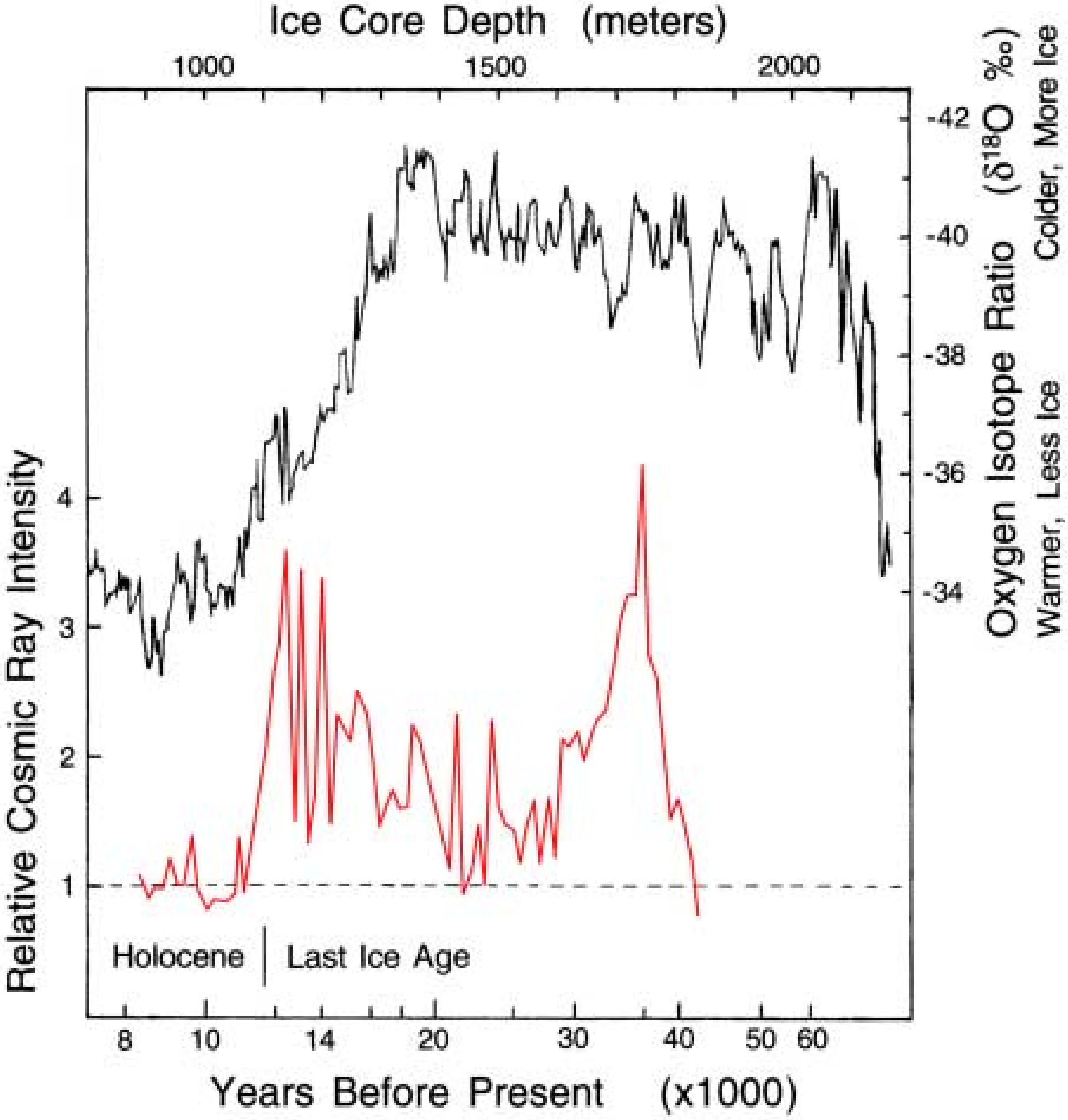

Figure 7. Lower profile: Cosmic ray intensity impacting the solar system (0 – 40 kyrs B.P.) normalized to present levels. (Based on the Byrd Station ice core $^{10}Be$ concentration data of Beer et al. (1987, p. 204; 1992, p. 145) adjusted for changes in ice accumulation rate and solar wind screening; see Appendix A, Part B.) Upper profile: The ice core's oxygen isotope ratio, and indicator of ambient temperature and glacial ice sheet size (courtesy of W. Dansgaard).

A variety of evidence indicates that the core of our Galaxy (Sgr A*), which lies 23,000 light years away, releases intense volleys of relativistic electrons about every $10^4$ years or so, and that these fronts, or *galactic superwaves*, travel radially outward through the Galaxy with such minimal dispersion that at the time of their passage they are able to elevate the cosmic ray background energy density in the solar neighborhood as much as $10^2$ to $10^3$ fold above current levels. During such a superwave passage, Galactic cosmic ray electrons would become trapped in spiral orbits behind the bow shock front that surrounds the heliopause and would develop energy densities $10^5$ fold higher than in interstellar space, reaching as high as $10^{-4}$ ergs/cm$^3$ and producing temperatures high enough to vaporize frozen cometary debris that currently orbits the solar system. Galactic superwaves are sufficiently intense and prolonged that they would propel the resulting interstellar/nebular dust and gas into the solar system which would have had a substantial effect on the Earth-Sun climate system. This theory and Galactic evidence that one such cosmic ray volley passed our solar system near the end of the last ice age is given in detail in other publications (LaViolette, 1983a, 1987, 2003).

Intense cosmic ray pulses from less massive stellar sources in the Galaxy such as Cygnus X-3 and Hercules X-1 are also known to maintain nondispersed configurations over distances of thousands of light years. For example, cosmic rays showering the Earth from Hercules X-1, which lies about 16,000 light years away, are known to cause a slight variation in the cosmic ray background intensity at 1.2357 second intervals in phase with the synchrotron radiation pulses from that source (Schwarzschild, 1988; Dingus et al., 1988; Lamb et al., 1988; Resvanis et al., 1988). However, the cosmic ray showers arriving from these stellar sources are relatively minor when compared with the intensities that periodically radiate from the Galactic center.

The recurrent $^{10}$Be peaks found in polar ice may record times when fronts of Galactic cosmic ray electrons were passing through the solar vicinity. Currently, $^{10}$Be is produced in the Earth's atmosphere almost entirely by cosmic ray protons, cosmic ray electrons currently making up only one percent of the total cosmic ray background. Moreover cosmic ray electrons are rather inefficient producers of $^{10}$Be, their main means of production being mainly via high energy gamma ray secondaries generated during their passage through the atmosphere which would in turn have a comparatively small cross section for $^{10}$Be production. So peaks that show a doubling or tripling of $^{10}$Be above background levels could very well reflect a hundred fold rise of Galactic cosmic ray electron intensities above the proton background level.

A cosmic ray-climate connection of this sort could explain why $^{10}$Be peaks occur at many of the major Pleistocene climatic boundaries, such as the particularly large peak that coincides with the Termination II boundary (Stage 5e/6) or the peaks that coincide with the transition from the Eemian interglacial to the semiglaciated Sangamon (Stage 5d/5e). A moderately high $^{10}$Be peak is also apparent in the Vostok profile at the Termination I boundary at around 13.7 kyrs B.P., confirming the prediction made earlier by LaViolette that a relatively intense cosmic ray volley passed through the solar system about 16 to 12 thousand years ago (1983a, 1987). This terminal $^{10}$Be event is better resolved in the Byrd Station $^{10}$Be production rate profile (Figure 7).

$^{10}$Be peaks are notably absent from the present interglacial, a period that appears to be unique for its long period of uniform climate. On the other hand, compared with this interglacial mean, the numerous peaks occurring during the last ice age (2 - 4), raised the $^{10}$Be production rate mean 50 percent higher and the peaks appearing during the semiglaciated Sangamon (5a - 5d) raised the $^{10}$Be mean 40 percent higher. So there appears to be a long-term correlation between climate and $^{10}$Be production rate (i.e., $^{10}$Be adjusted for changes in accumulation rate). The previous interglacial (Stage 5e) also seems to have been free of $^{10}$Be peaks.

Supernova explosions cannot reasonably account for the recurrent $^{10}$Be peaks evident in the polar ice record since sufficiently close supernovae occur very rarely, only about once every $10^8$ years in the solar neighborhood. Nevertheless, Konstantinov, et al. (1990) and Sonett (1991) have proposed that $^{10}$Be peaks dating at around 35 and 60 kyrs B.P. may have been produced by cosmic ray blast waves arriving from s nearby supernova explosion. Konstantinov et al. suggest that it was located about 180 ± 20 light years away. Sonett associates it with the explosion that formed the North Polar Spur and cites an explosion date of 75 kyrs B.P. (Davelaar et al., 1980) which presumes that the remnant achieved its rather large size (~370 light years) as a result of an unusually energetic explosion occurring in a very rarefied interstellar medium. However, others believe that this remnant is of a much older age, about $10^6$ yrs (Heiles, 1980). Its slow rate of expansion, presently 3 km/sec, and other evidence, suggest that the it is instead a very old reheated remnant that arose from a supernova explosion of average energy release occurring in a region of normal interstellar gas density (Borken and Iwan, 1977). If this older age is valid, the North Polar Spur cosmic ray blast wave would have passed Earth hundreds of thousands of years earlier and hence its $^{10}$Be signature would not be registered in the polar ice record.

<u>Interstellar Dust Incursion</u>

There is plenty of frozen material both in and around the solar system which could be vaporized and propelled into the inner solar system by a Galactic superwave. Observations of infrared excesses in nearby stars suggest that the solar system, like these other star systems, is surrounded by a light-absorbing dust shell, and may contain about $10^3$ times more dust than had been previously supposed on the basis of IRAS observations of the zodiacal dust cloud (Aumann, 1988). Other observations indicate that the Sun is presently passing through an interstellar cloud that appears to be a component of the outer shell of the North Polar Spur supernova remnant, the closest supernova remnant to the Sun (Frisch, 1981; Frisch and York, 1983). So it is quite likely that the solar system has acquired this dust relatively recently, e.g., within the past several million years. This ongoing encounter may be responsible for the long-period comets that periodically enter the solar system from directions within 5° - 10° of the solar apex, the direction of the Sun's motion through the interstellar medium. Since the solar apex changes its orientation by about 1.5° per Myrs due to the Sun's motion around the Galactic center, it may be surmised that the Sun acquired these comets sometime within the past 3 to 6 million years (Clube and Napier, 1984). This is comparable to the time span of the present glacial cycle sequence.

This proximal remnant may also be the source of the billion or more cometary masses estimated to be present in the Edgeworth-Kuiper belt that begins just beyond the orbit of Neptune and extends outward a hundred AU or more (Horgan, 1995). In addition, Ulysses spacecraft observations have shown that an ecliptic ring of dust is present whose inner edge begins just outside the orbit of Saturn and which contains dust at a density $10^4$ times higher than in the vicinity of the Earth (Landgraf, 2002). Furthermore this dusty environment may explain why interstellar dust grains are currently entering the solar system and dominating the dust particle population outside the asteroid belt (Grün et al., 1993).

This influx of interstellar dust would explain the alignment of the zodiacal dust cloud's ecliptic nodes. In 1984, the IRAS (InfraRed Astronomy Satellite) team noted that their observations confirmed earlier reports that zodiacal cloud is tilted about 3 degrees relative to the ecliptic with a descending ecliptic node at ecliptic longitude $\lambda = 267 \pm 4°$ (Hauser et al., 1984). LaViolette (1987) concluded that the proximity of this nodal alignment to the Galactic center direction could be explained if dust forming the outer zodiacal cloud was of interstellar origin and had recently entered the solar system from the Galactic center direction. He noted that this confirmed his earlier prediction that interstellar dust should have recently entered the solar system driven by a cosmic ray wind emanating from the Galactic center direction (LaViolette, 1983a). In 1998, the Diffuse Infrared Background Experiment (DIRBE) team more accurately located the position of the zodiacal cloud's descending ecliptic node to lie at $\lambda = 257.7 \pm 0.6°$ (Kelsall, et al., 1998). In galactic coordinates this is positioned at ($\ell$ = 0.5± 0.6°, b = +10 ± 0.6°) and coincides with the Galaxy's zero longitude meridian; see point A in Figure 8.

The 1993 Ulysses data which showed that a flux of interstellar dust was currently entering the solar system from the Galactic center direction confirmed the earlier prediction of LaViolette (1993) of recent interstellar dust entry based on the zodiacal ecliptic node position. In fact, (Witte, et al., 1993) reported that this dust influx was entering from the same direction as the 26 km $sec^{-1}$ interstellar helium wind which they observed coming from the galactic direction ($\ell$ = -0.2 ± 0.5°, b = +16.3 ± 0.5°). Not only does this direction lie within a few degrees of the ecliptic, it also coincides with the Galaxy's zero longitude meridian as does the zodiacal dust cloud node; see point B in Figure 8. These interstellar gas and dust winds may be relics of more intense influxes driven by past cosmic ray superwaves emanating from the Galactic center (LaViolette, 1983a, 1987). The incident cosmic rays would magnetically couple a portion of their kinetic energy to electrically charged dust particles and gas ions driving this material forward. This same residual Galactic cosmic ray wind could explain why $H_3^+$ ions have been detected in nearby diffuse interstellar clouds at concentrations much higher than expected (McCall, 2003).

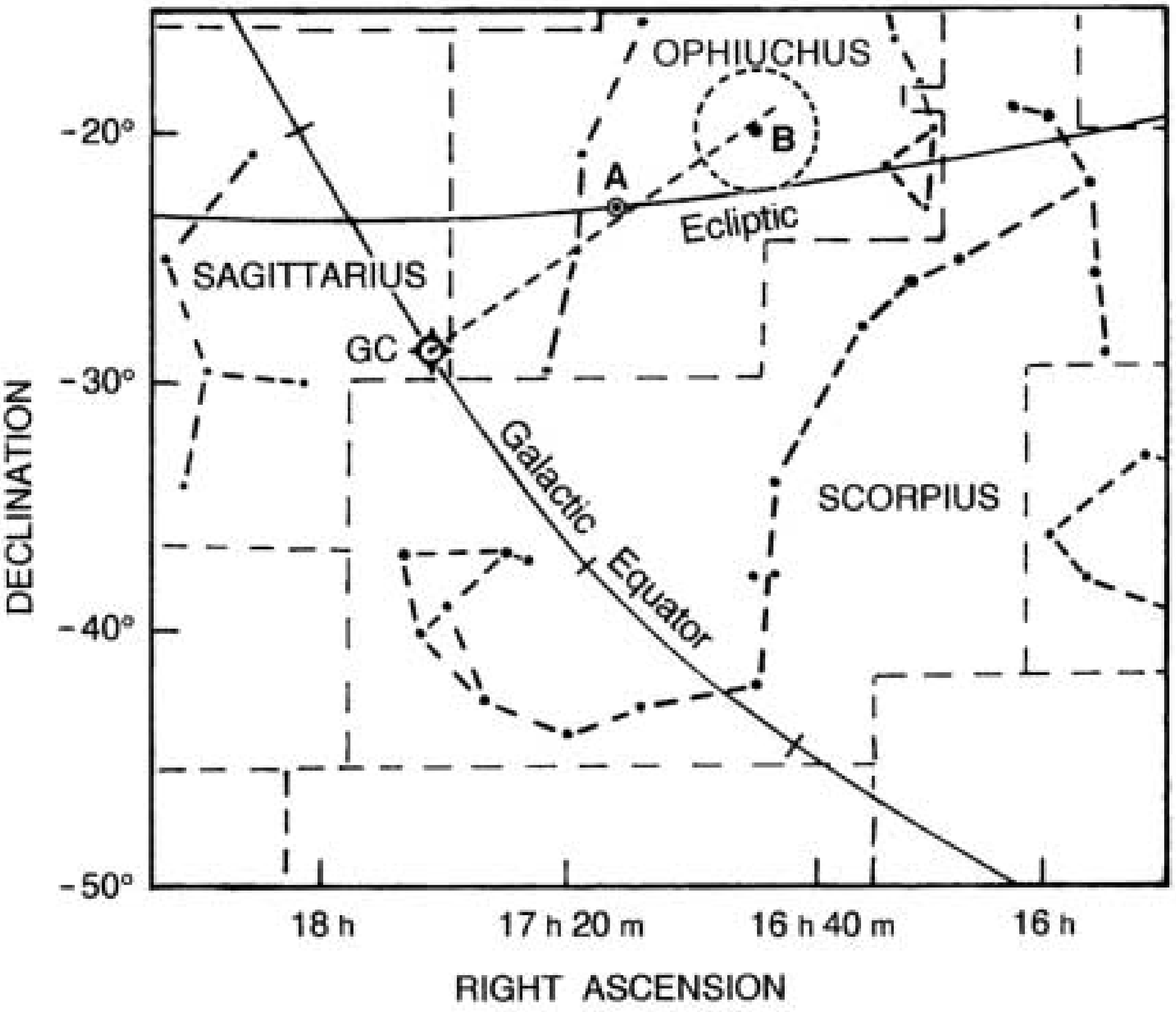

Figure 8. Sky map of the Scorpius region. GC marks the location of the Galactic center. Point A indicates the ascending node that marks the intersection of the zodiacal dust cloud orbital plane with the ecliptic plane. Point B indicates the direction from which the interstellar helium wind and interstellar dust particle wind are entering the solar system.

According to Ulysses measurements, the interstellar dust particles presently entering the solar system span the mass range $10^{-15}$ - 5 X $10^{-12}$ g, or a size range 0.1 - 1.5 mm, assuming a particle mass density of $r$ ~ 3 g $cm^{-3}$. Unlike dust in the Earth's immediate vicinity, whose size distribution peaks at 200 to 400 mm (Hughes, 1975), particles in this size range very effectively scatter and absorb solar radiation. So the possibility that large quantities of such interstellar dust may have entered the solar system in Earth's recent past should be a matter of concern from a climatological standpoint.

Evidence of Extraterrestrial Dust in the Polar Ice Record.

The unusually high concentrations of HF and HCl acids found in Byrd Station, Antarctic ice dating about 15,800 year B.P., may be residues from one such interstellar dust incursion. Hammer, et al. (1997) note that it is difficult to explain these eight peaks as having a volcanic origin because the combined acid output which spans a period of about a century exceeds by 18 fold the largest volcanic signal observed in the Byrd ice core record and also because the recurrence of the events is unusually regular, a behavior that is not seen in volcanic eruptions. It has been shown that the peaks recur with an average period of $11.5 \pm 2.4$ years which matches the solar cycle period indicating that the deposited acids and their associated dust may be of interstellar origin (LaViolette (2005). That is, the influx of interstellar dust would similarly vary in intensity according to the solar cycle period since its entry is modulated by the regularly changing orientation of the Sun's magnetic field. The entry of this material may have been associated with a Galactic superwave since, as seen in Figures 6 and 7, this 15,800 year acidity event coincided with a rise in the cosmic ray background intensity, an increased that persisted with some variation until the end of the ice age.

It is significant that this so called "Main Event" falls at the beginning of the Pleistocene deglaciation. The deposition of this acid bearing dust was initially punctuated by an abrupt climatic cooling which is registered in both the Greenland and Byrd Station, Antarctic ice records, and was followed by the 900 year long Pre-Bölling interstadial warming. After a brief stadial this was followed by the Allerod-Bölling interstadial sequence. So, the discovery of a possible extraterrestrial origin of the Main Event leads suggests that the global warming which followed may have been extraterrestrially induced.

It is estimated that the deposition rate of this dust, of $1.4 \times 10^{-9}$ g/m$^2$/s, would project dust concentrations in the vicinity of the Earth of $5 \times 10^{-20}$ g/cm$^3$ which could have presented an optical depth of up to 0.2 between the Earth and the Sun (LaViolette, 2005). The present concentration of interplanetary dust in the vicinity of the Earth is estimated to be about $2 \times 10^{-22}$ g cm$^{-3,}$ of which only about 0.02 percent consists of particles in the 0.2 mm size range having a maximal cross–section for absorbing or scattering sunlight. If the invading interstellar dust particles were of submicron size similar to those observed by Ulysses, the influx would have increased the concentration of optically interactive particles in the solar system by over a million fold. The initial climatic cooling effect at the time of the event could have been brought about by the prolonged presence of light scattering interstellar dust particles in the Earth's stratosphere. The subsequent deglacial warming could have been due to a combination of factors: a) destruction of the ozone layer due to the presence of interstellar halides allowing UV penetration, b) increase of the solar constant due to light backscattered from the zodiacal dust cloud, c) shift of the incident solar spectrum to the infrared resulting in greater absorption of the solar beam (reduced scattering from high albedo surfaces), and d) a major increase in the Sun's luminosity and activation of its photosphere and corona due to the dust's effect on the Sun (LaViolette, 1983a, 2005). There is evidence that solar flare activity was one to two orders of magnitude higher during this deglacial period (Zook, et al., 1977; Zook, 1980).

Future work should establish the possible extraterrestrial origin of the Main Event by searching this ice core horizon for the presence of cosmic dust indicators. However, a previous study has found elevated levels of cosmic dust in Camp Century, Greenland polar ice. Samples ranging from 34 to 70 thousand years old were found to contain high concentrations of iridium and nickel in proportions similar to those found in extraterrestrial material; see Figure 9, lower profile (LaViolette, 1983a, 1983b, 1985).[2] Some of these ice age samples had Ir deposition rates $10^3$ times greater than those reported by Takahashi et al. (1978) for recent Camp Century snows, implying that at certain times during the Wisconsin cosmic dust deposition rates were substantially higher than at present. Assuming that the Ir in the Wisconsin stage ice came from an interstellar dust source having an Ir composition similar to that found in carbonaceous chondrites, these measurements project an interstellar dust influx rate of up to $3 \times 10^{-6}$ g cm$^{-2}$ yr$^{-1}$, implying that the near-Earth interplanetary dust concentration may have reached as high as $3 \times 10^{-20}$ g cm$^{-3}$, 150 fold higher than its present value. Including the unaccounted for fraction present as water would bring the concentration to a level comparable with what is projected for the Main Event dust incursion.

Figure 9 compares the Camp Century Ir deposition rate data to the $^{10}$Be concentration profile of Beer et al. (1992). Due to gaps in the $^{10}$Be data, it is possible to correlate only the youngest of the eight cosmic dust values (1215.1 m, ~34 k yrs B.P.) with a $^{10}$Be data point. This sample, which had the second highest Ir concentration of the eight samples, coincides with the latter part of a major $^{10}$Be peak dated at the 35 kyr B.P. Although this correspondence is consistent with the proposed cosmic-ray/cosmic-dust causal link, additional measurements of both Ir and $^{10}$Be in ice age polar ice are needed in order to decide whether there is a clear connection.

## 5. Conclusion

Taken as a whole, the available paleoclimatological data suggest that, during the Termination I deglaciation, temperatures in many locations around the world underwent coordinated changes, with major warmings occurring during the Bölling-Allerod (14.5 k - 12.7 k cal yrs B.P.) and Preboreal (11.55 k - 11.3 k cal yrs B.P.) and with a more minor interstadial centered around 15.3 k cal yrs B.P. Available data suggests that these warmings were initiated neither by changes in atmospheric $CO_2$ concentration nor by a major alteration in the rate of North Atlantic deep-water production. Moreover it is not clear whether these mechanisms are capable of producing warmings and coolings of the kind of magnitude, geographical extent, and abruptness observed at the Termination I boundary. Polar ocean front migrations and weather fluctuations also do not offer an adequate explanation.

---

[2] These ice core sample ages given here are older than reported in the original publications. The new ages reflect a subsequent revision of the Camp Century ice core chronology after it had become keyed to the Summit, Greenland chronology.

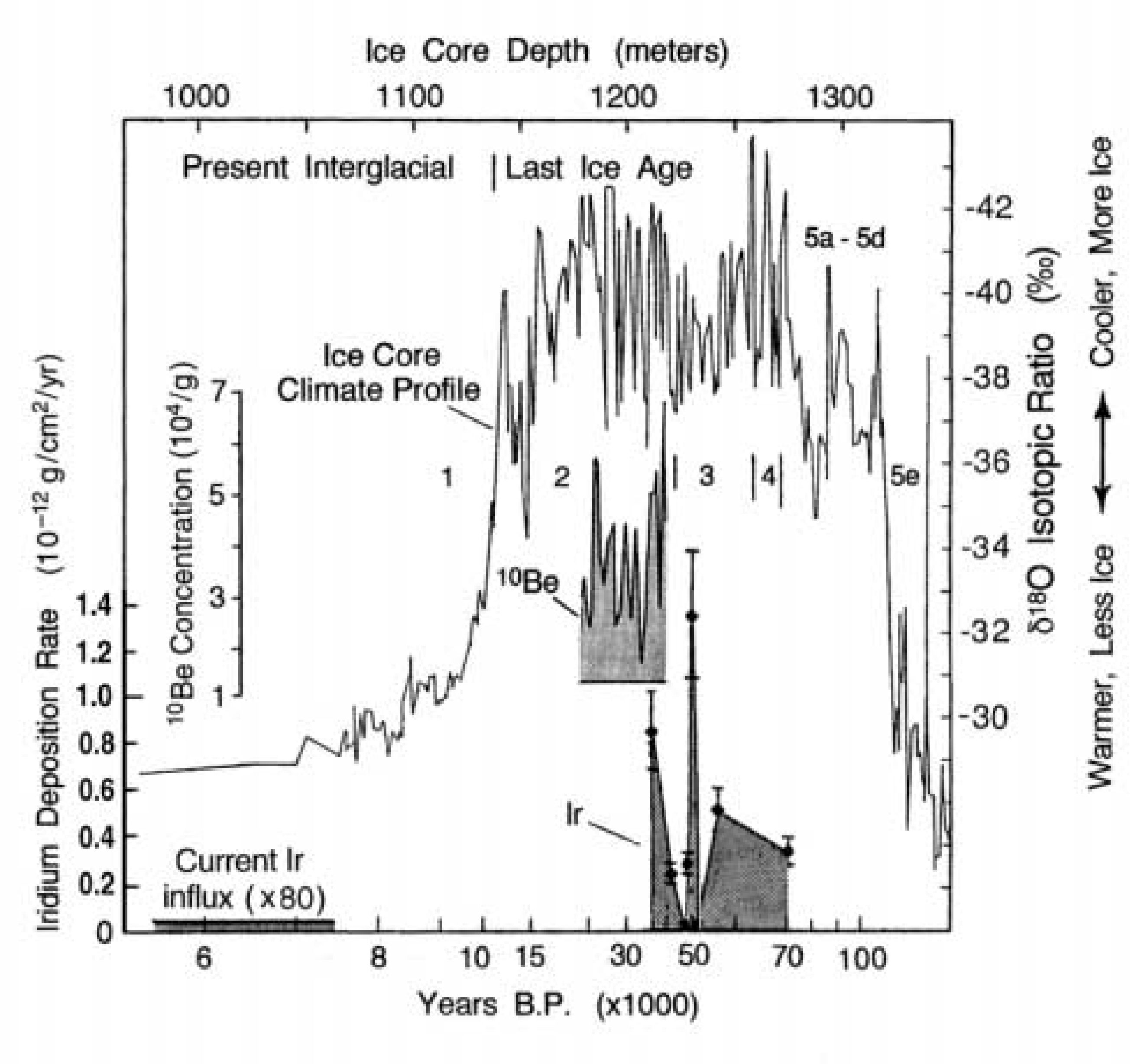

Figure 9. Upper curve: Camp Century ice core oxygen isotope climatic profile (Dansgaard, personal communication, 1982). Middle curve: cosmogenic beryllium concentration (after Beer et al., 1992). Lower curve: iridium deposition rate for the Holocene (Takahashi, 1978) and the Wisconsin (LaViolette, 1983a, 1985), assuming ice accumulation rates of 38 and 15 cm $yr^{-1}$ respectively.

Evidence that the solar system resides in a dust congested environs, of a current influx of interstellar dust, of acid residues in 15,800 year old polar ice bearing a solar cycle signature, of episodes of accelerated deposition of cosmogenic beryllium, Ir, and Ni during the Pleistocene, and of intense solar flare activity at the end of the ice age together suggest that the Termination I deglaciation, and other climatic transitions before it, may have been extraterrestrially induced. Astronomical evidence suggests that intense volleys of Galactic cosmic rays periodically pass through the solar vicinity from the direction of the Galactic center, the most recent volley passing through toward the end of the last ice age. These prolonged cosmic ray assaults would have propelled interstellar dust and gas into the solar system at rates much higher than currently observed rates. This material could have activated the Sun, altered the intensity and spectrum of its radiation, and changed the Earth's stratospheric albedo. Depending on the relative weightings of these effects, this could have led either to rapid surface cooling and ice sheet advance or to rapid surface warming and ice sheet recession.

Such extraterrestrial disturbances could account for the abruptness and global coherence of climatic transitions observed in the terrestrial record. Moreover such short-period stochastic forcings could account for a large percentage of the variance in the Earth's ice volume record which is not explained by orbital cycle forcing. The intensity of these external perturbations and the prevailing terrestrial boundary conditions (e.g., ice sheet size, orbital parameter phase, atmospheric $CO_2$ concentration, and deep-water production rate) would together determine whether climate became either temporarily perturbed or flipped into a long-term glacial or interglacial mode. In particular, these terrestrial factors in combination could explain why the climatic system became stabilized in an interglacial mode following the Termination-I warming events.

Future ice core measurements charting the temporal variation of cosmic dust concentrations and their correlation with $^{10}$Be and stable isotope variations should help to elucidate the connection between Galactic cosmic ray intensity, cosmic dust influx rate, solar activity, and climate.

**Acknowledgments**

I would like to thank Fred LaViolette, George Lendaris, and others for helpful comments on this manuscript. I would also like to thank J. Jouzel, S. Johnsen, J. Kennett, J. Beer, and M. P. Ledru for sharing their deuterium, $\delta^{18}$O, $^{10}$Be, and pollen count data which proved to be of great help.

## APPENDIX A

A. Adjustment of the Vostok $^{10}Be$ Data.

1. Ice Accumulation Rate Adjustment

The relative cosmic ray intensity profile shown in Figure 6 was obtained by converting the $^{10}Be$ concentration values (*C* atoms/g) of Raisbeck et al. (1987) into normalized $^{10}Be$ production rates ($\Phi$ atoms/cm$^2$/yr) according to the formula:

$$\Phi = (C \bullet a \bullet \rho)/k, \qquad \text{(A-1)}$$

where $a$ is the variable ice accumulation rate at Vostok given in Column 6 of Table A-I, $\rho = 0.917$ g/cm$^3$ is the density of ice, and $k = 1.75 \times 10^5$ atoms/cm$^2$/yr is the Holocene $^{10}Be$ production rate average used to normalize the product. For values spanning the period 15,500 to 10,900 years B.P. the values are boosted by the values given in column 4 of Table A-II to adjust for the increased solar screening due to the increased solar activity prevailing during that time (see No. 2 of this section below).

Ice accumulation rates for Vostok (Table A-I, Column 6) have been estimated throughout the core by the formula $a = \lambda \bullet \tau$, where $\lambda$ is the estimated annual layer thickness of the ice (Column 5) calculated using calendar dates assigned to various core depths (Columns 1 & 3) and where $\tau$ is the correction for plastic deformation of the ice sheet (Column 4). The correction for plastic deformation is calculated according to the linear relation $\tau = 0.96 \bullet 3700/(3700 - d)$, where 3700 is the present meter thickness of the ice sheet, d is the sample depth in meters, and 0.96 is an adjustment factor reflecting the assumption that the ice sheet was thicker during the last ice age. For the Holocene (0 - 280 m depth), 15 meters have been subtracted to figure the average accumulation rate (i.e., 269m/11,580 years). After correcting for deformation, this yields a = 2.4 cm/yr, which agrees with the present ice accumulation rate at Vostok.

Calendar dates listed in Column 1 were assigned to the Vostok ice core by correlating specific climatic features in the Vostok deuterium profile with similar features in dated climatic profiles. The climatic boundaries between depths 354 m to 284 m are dated based on correlations to the GRIP ice core chronology; see Table I for dates. The 29, 50, 64, and 68 kyrs B.P. $^{14}C$ dates are cited from Woillard and Mook (1982) and Grootes (1978). Calendar dates prior to 30 kyrs B.P. are consistent with the accepted U/Th dates for the respective climatic boundaries.

2. The Solar Modulation Adjustment

Solar flare activity has an inverse effect on terrestrial cosmic ray intensity, decreasing cosmic ray intensities at times of high solar activity by increasing the effect of solar wind screening. To determine cosmic ray intensities outside the solar system, we must adjust for solar wind screening at times when there is reason to believe that the Sun was particularly active. $^{10}Be$ production rate, which is inferred from polar ice core data, is an indicator of the terrestrial cosmic ray flux. Consequently, during periods of high solar activity, the $^{10}$Be production rates must be proportionately inflated to indicate cosmic ray levels prior to solar screening effects. Available data indicates that the Sun was unusually active during the global warming period at the end of the last ice age from about 16,000 to 11,000 years BP. It is likely that the Sun was also particularly active at earlier times, particularly during interstadial periods (e.g., 36 - 31 kyrs BP) and during the termination of the previous ice age (136 - 128 kyrs BP). However since data is lacking on the degree of solar activity during these periods, the data has been adjusted only for the period ending the last ice age.

Table A-I
Chronology, Accumulation Rate Adjustments, and
Climatic Zone Correlations for the Vostok Ice Core

| (1) | (2) | (3) | (4) | (5) | (6) | (7) | (8) | (9) |
|---|---|---|---|---|---|---|---|---|
| YEARS B.P. absol. | C-14 | DEPTH (m) | Deform Correc. | Ann. cm $yr^{-1}$ | Accum. Rate | CLIMATIC Europe | PHASE N. America | CLIMATIC BOUNDARIES |
| 0 | 0 | | | | | | | |
| | | | 1.04 | 2.33 | 2.42 | | | |
| 11.55 | 10.0 | 284 | | | | Y. Dryas ends | | H / LW |
| | | | 1.042 | 1.39 | 1.45 | | | |
| 12.7 | 11.0 | 300 | | | | Y. Dryas begins | | 1/2 |
| | | | 1.048 | 1.64 | 1.71 | | | |
| 13.25 | 12.0 | 309 | | | | IntraAllerod cold peak begins | | |
| | | | 1.052 | 2.08 | 2.19 | | | |
| 14.5 | 13.0 | 335 | | | | Bölling begins | Cary/Port Huron Inter. begins | |
| | | | 1.056 | 1.43 | 1.51 | | | |
| 14.85 | 13.3 | 340 | | | | Pre-Bölling Inter. begins | | |
| | | | 1.058 | 1.56 | 1.65 | | | |
| 15.75 | 14.2 | 354 | | | | Pre-Bölling Inter. begins | | |
| | | | 1.096 | 1.32 | 1.45 | | | |
| 32.0 | 29.0 | 570 | | | | Denekamp Inter. | Plum Point Inter. | 2/3, LW / MW |
| | | | 1.15 | 1.42 | 1.63 | | | |
| 38.0 | | 655 | | | | | Port Talbot–2 Inter. | |
| | | | 1.21 | 1.22 | 1.48 | | | |
| 54.0 | 50.0 | 850 | | | | Moershoofd Inter. | Port Talbot–1 Inter. | |
| | | | 1.26 | 1.26 | 1.59 | | | |
| | | 920 | | | | | | 3/4, MW / EW |
| | | | 1.30 | 1.26 | 1.64 | | | |
| 67.5 | 64.0 | 1020 | | | | Börup Inter. | St. Pierre Inter. | |
| | | | 1.33 | 1.2 | 1.60 | Nicolet Stad. | | 4/5, EW / S |
| 70.0 | 68.0 | 1050 | | | | Amersfoort Inter. | | |
| | | | 1.50 | 1.43 | 2.15 | | | |
| 110 | | 1620 | | | | | | 5e/d |
| | | | 1.78 | 1.44 | 2.57 | | | |
| 122.5 | | 1800 | | | | | | |
| | | | 1.91 | 1.45 | 2.77 | | | |
| 128 | | 1880 | | | | | | 5/6, S / I |
| | | | 1.99 | 1.30 | 2.58 | | | |
| 133 | | 1945 | | | | | | |
| | | | 2.08 | 0.83 | 1.73 | | | |
| 148 | | 2070 | | | | | | |

The adjustments were made as follows. Hughen et al. (1998) have measured radiocarbon anomalies for the period from 14,200 to 9,000 calendar years B.P. Based on their data, Column 3 of Table A-II gives the percent change in the atmospheric radiocarbon concentration. Here we adopt a baseline that is normalized to their Holocene data and which is 1.5% lower than the zero reference point given by Hughen et al. To show how the adjustment is carried out, let us take as an example the data point around 12,600 years BP when atmospheric radiocarbon reached a maximum level of 9.5% above normal. Beer et al. (1985) note that the variation in $^{14}C$ concentration induced by the 11-year solar cycle is attenuated by a factor of 100 because of the rapid transfer of $^{14}C$ from the atmosphere to the geosphere. Since solar cycle variations typically produce a 0.3% change in atmospheric $^{14}C$ concentration, a 9.5 percent peak increase in atmospheric$^{14}C$ would translate into a 30 fold increase in solar cosmic ray activity (9.5/0.3) if it had occurred over a comparable 11 year time period. Since the anomaly was sustained over centuries rather than over a decade, we will assume that solar flare cosmic ray flux underwent only a 2.43 fold peak increase, calculated by inflating the percent rise in atmospheric carbon by a factor of 15; i.e., (0.095 X 15) + 1 = 2.43. Column 4 of Table A-II charts this inferred change in solar flare cosmic ray intensity based on this $^{14}C$ data listed in Column 3. Note that the inferred increase in solar cosmic ray intensity is much less than the increase implied by the moon rock data of Zook et al. (1977, 1980) (see Column 2 of Table A-II). So, this rough estimate appears to be on the conservative side.

Since the Galactic and extragalactic cosmic ray intensity incident on the solar system is screened by an amount that is proportionate to the level of solar cosmic ray intensity, a 2.43 fold increase in solar cosmic ray intensity would cause a proportionate decrease in background flux reaching the Earth. So to correct for this decrease we must boost the calculated $^{10}Be$ production rate data by 2.43 fold. to compensate for the increased solar wind screening.

Table A-II
Determining the Galactic Cosmic Ray Intensity Adjustment Factor

| 1 | 2 | 3 | 4 |
|---|---|---|---|
| Years BP | Normalized solar cosmic ray intensity (Zook) | Δ C-14 (percent) | Normalized solar cosmic ray intensity |
| 10.0 - 10.9 | 8 | 0 | 1.00 |
| 10.9 - 11.1 | 10 | 4.0 | 1.60 |
| 11.1 - 11.8 | 11 | 1.0 | 1.15 |
| 11.8 - 12.5 | 14 | 4.0 | 1.60 |
| 12.5 - 12.7 | 15 | 9.5 | 2.43 |
| 12.7 - 12.9 | 16 | 6.5 | 1.98 |
| 12.9 - 13.1 | 17 | 6.0 | 1.90 |
| 13.1 - 13.7 | 20 | 2.5 | 1.38 |
| 13.7 -13.9 | 21 | 5.0 | 1.75 |
| 13.9 - 14.7 | 25 | 2.5 | 1.38 |
| 14.7 - 15.1 | 27 | 1.5 | 1.23 |
| 15.1 - 15.5 | 35 | 1.0 | 1.15 |
| 15.5 - 16.0 | 50 | 0.5 | 1.08 |

B. <u>Ice Accumulation Rate Adjustment of the Byrd Station, Antarctica $^{10}$Be Data.</u>

The $^{10}$Be profile shown in Figure 7 was obtained by converting the $^{10}$Be concentration values ($C$ atoms/g) of Beer et al. (1992) into normalized $^{10}$Be production rates ($\Phi$ atoms/cm$^2$/yr) according to formula (1), where $a$ is the variable ice accumulation rate at Byrd Station given in Column 7 of Table A-III, and k = 1.59 X 10$^5$ atoms/cm$^2$/yr is the Holocene $^{10}$Be production rate average used to normalize the product. For the period 15,500 to 10,900 years B.P., the projected relative cosmic ray intensity values have been boosted by the amounts given in Column 4 of Table A-II to adjust for the increased solar screening due to the increased solar activity prevailing during that time; see No. 2 of Section A above.

Ice accumulation rates for Byrd Station (Table A-III, Column 7) have been estimated throughout the core in the same fashion as for Vostok by the formula $a = \lambda \bullet \tau$, where $\lambda$ (Column 5) is the estimated annual layer thickness of the ice (cm/year) calculated from the calendar dates

Table A-III

Chronology, Accumulation Rate Adjustments, and
Climatic Zone Correlations for the Byrd Station Ice Core

| (1) | (2) | (3) | (4) | (5) | (6) | (7) | (8) |
|---|---|---|---|---|---|---|---|
| YEARS B.P. absol. | C-14 | GRIP Depth (m) | BYRD Depth (m) | $\lambda$ (cm yr$^{-1}$) | Deform Correc. $\tau$ | Accum. Rate (cm yr$^{-1}$) | Climatic boundary |
| 0 | 0 | | | | | | |
| | | | | 9.26 | 1.34 | 12.4 | |
| 11.55 | 10.0 | 1623 | 1100 | | | | Younger Dryas Stad. ends |
| | | | | 3.74 | 1.99 | 7.4 | |
| 12.7 | 11.0 | 1663 | 1143 | | | | Younger Dryas Stad. begins |
| | | | | 4.87 | 2.09 | 10.2 | |
| 13.87 | 12.0 | 1718 | 1200 | | | | Older Dryas / Allerod begins |
| | | | | 4.60 | 2.17 | 9.9 | |
| 14.5 | 13.0 | 1754 | 1229 | | | | Bölling Inter. begins |
| | | | | 3.14 | 2.21 | 6.9 | |
| 14.85 | 13.3 | 1766 | 1240 | | | | Lista Stad. begins |
| | | | | 4.40 | 2.27 | 10.0 | |
| 15.75 | 14.2 | 1795 | 1280 | | | | Pre-Bölling Inter. begins |
| | | | | 2.67 | 2.66 | 7.1 | |
| 24.0 | | | 1500 | | | | |
| | | | | 2.25 | 3.47 | 7.8 | |
| 32.0 | 30.0 | 2177 | 1680 | | | | Denekamp Inter. |
| | | | | 2.33 | 4.23 | 9.9 | |
| 35.0 | | | 1750 | | | | Beryllium-10 marker peak |
| | | | | 1.32 | 4.99 | 6.6 | |
| | | | 1840 | | | | Port Talbot-2 Inter. |
| | | | | 1.32 | 7.25 | 9.6 | |
| 54.0 | 50.0 | | 2000 | | | | Moershoofd / Port Talbot-1 Inter. |
| | | | | 0.59 | 11.1 | 6.6 | |
| 67.5 | 64.0 | | 2080 | | | | Börup / St. Pierre Inter. |

(column 1) that have been assigned to various core depths (column 4) and where $\tau$ is the correction for plastic deformation of the ice sheet (column 6). The deformation correction is calculated according to the linear relation $\tau = 2250/(2250 - d)$, where 2250 is the height of the ice sheet on the assumption that during the last ice age the ice sheet was 4 per cent thicker than it is at present, and d is the sample depth. For the Holocene (0 - 1100 m depth), this method gives an annual accumulation rate of 12.4 cm/yr, which agrees with the present ice accumulation rate at this location.

C. Ice Core Chronology and the Assumption of Synchronous Climatic Change

The above ice core chronologies are derived by correlating climatic boundaries seen in the Byrd and Vostok ice core oxygen isotope profiles with those seen in the well-dated GRIP ice core from Summit, Greenland (Johnsen, et al., 1992). In correlating the ice core isotope profiles, we have assumed that major changes in climate occur contemporaneously in both the northern and southern hemispheres and hence that distinct climatic change boundaries evident in the GRIP ice core may be matched up with similar boundaries in the Byrd Station and Vostok ice cores. The assumption that the Earth's climate warmed and cooled in a globally synchronous manner at the end of the last ice age is supported by evidence from dated land, sea, and ice climate profiles which show that the Bölling/Alleröd/Younger Dryas oscillation occurred synchronously in both northern and southern latitudes. This evidence has been reviewed above in Section 2. The chronology adopted here for the Byrd core is consistent with that of Beer et al. (1992) which was obtained by correlating distinctive $^{10}Be$ concentration peaks found in both the Byrd Station, Antarctica and Camp Century, Greenland isotope records, some peaks dating as early as 12 – 20 kyrs BP. The Camp Century isotope profile, in turn, has been accurately dated through correlation with the annual layer dated Summit, Greenland isotope profile.

## References

Alley, R.B. et al., 1993 Abrupt increase in Greenland snow accumulation at the end of the Younger Dryas event. Nature, 362: 527-529.

Atkinson, T.C., Briffa, K.R. and Coope, G.R., 1987. Seasonal temperatures in Britain during the past 22,000 years, reconstructed using beetle remains. Nature, 325: 587-592.

Aumann, H.H., 1988. Spectral class distribution of circumstellar material in main-sequence stars. A.J., 96: 1415-1419.

Bard, E., Fairbanks, R., Arnold, M., Maurice, P., Duprat, J., Moyes, J. and Duplessy, J.-C., 1989. Sea-level estimates during the Last deglaciation based on $\delta^{18}$O and accelerator mass spectrometry $^{14}$C ages measured in *Globigerina bulloides*. Quat. Res., 31: 381-391.

Bard, E., Hamelin, B., Fairbanks, R.G. and Zindler, A., 1990a. Calibration of the $^{14}$C timescale over the past 30,000 years using mass spectrometric U-Th ages from Barbados corals. Nature, 345: 405-410.

Bard, E., Hamelin, B. and Fairbanks, R.G., 1990b. U-Th ages obtained by mass spectrometry in corals from Barbados: sea level during the past 130,000 years. Nature, 346: 456-458.

Barnola, J.M., Raynaud, D., Korotkevich, Y.S. and Lorius, C., 1987. Vostok ice core provides 160,000-year record of atmospheric $CO_2$. Nature, 329: 408-414.

Beard, J.H., 1973. Pleistocene-Holocene boundary and Wisconsin substages in the Gulf of Mexico. In: R.F. Black, R.P. Goldthwait, and H.B. Willman (Editors) The Wisconsin Stage (GSA Memoir 136). GSA, Boulder, CO, pp. 277-297.

Beer, J., et al., 1984a. Temporal variations in the $^{10}$Be concentration levels found in the Dye-3 ice core, Greenland. Ann. Glaciol., 5: 16-17.

Beer, J, et al , 1984b. The Camp Century $^{10}$Be record: Implications for long-term variations of the geomagnetic dipole moment. Nuc. Instrum. Meth. B5: 380-384.

Beer, J., et al., 1985. $^{10}$Be Variations in polar ice cores. In: C.C. Langway, Jr., H. Oeschger, and W. Dansgaard (Editors) Geophysics, Geochemistry and the Environment (AGU Monograph No. 33). AGU, Washington, D.C., pp. 66-70.

Beer, J., et al., 1987. $^{10}$Be measurements on polar ice: Comparison of Arctic and Antarctic records. Nuclear Instruments and Methods in Physics Research, B29: 203−206.

Beer, J, Siegenthaler, U., Bonani, G., Finkel, R.C., Oeschger, H., Suter, M., and Wölfli, W., 1988. Information on past solar activity and geomagnetism from $^{10}$Be in the Camp Century ice core. Nature 331: 657-679.

Beer, J., et al., 1992. $^{10}$Be peaks as time markers in polar ice cores. In: The Last Deglaciation: Absolute and Radiocarbon Chronologies (Proc. NATO ASI Series, vol. 12). Springer-Verlag, Heidelberg, pp. 140-153.

Berger, W.H., 1990. The Younger Dryas cold spell – a quest for causes. Paleogeogr. Paleoclimatol. Paleoecol. (Global Plan. Change), 89: 219-237.

Berglund, B.E., 1979. The deglaciation of southern Sweden 13,500 - 10,000 B.P. Boreas, 8: 89-118.

Björck, S., and Möller, P., 1987. Late Weichselian environmental history in southeastern Sweden during the deglaciation of the Scandinavian ice sheet. Quat. Res., 28: 1-37.

Blunier, T. et al., 1998. Asynchrony of Antarctic and Greenland climate change during the last glacial period. Nature, 394: 739-743.

Borken, R.J., and Iwan, D.C., 1977. Spatial structure in the soft X-ray background as observed from OSO-8, and the North Polar Spur as a reheated supernova remnant. Ap.J., 218: 511-520.

Boyle, E.A., and Keigwin, L.D., 1987. North Atlantic thermohaline circulation during the past 20,000 years linked to high-latitude surface temperature. Nature, 330: 35-40.

Broecker, W.S., Peteet, M. and Rind, D., 1985. Does the ocean-atmosphere system have more than one stable mode of operation? Nature, 315: 21-25.
Broecker, W.S. et al., 1988a. The chronology of the last deglaciation, Implications to the cause of the Younger Dryas event. Paleoocean, 3: 1-19.
Broecker, W.S. et al., 1989. Routing of meltwater from the Laurentide Ice Sheet during the Younger Dryas cold episode. Nature, 341: 318-321.
Broecker, W.S., and Denton, G.H. 1990. What drives glacial cycles?. Sci. Am., 262(1): 49-56.
Burrows, C.J., 1979. A chronology for cool-climate episodes in the Southern Hemisphere 12,000 - 1000 yr. B.P. Paleogeogr. Paleoclimatol. Paleoecol., 27: 287-347.
Chappellaz, J., Blunier, T., Raynaud, D., Barnola, J. M., Schwander, J., and Stauffer, B., 1993. Synchronous changes in atmospheric $CH_4$ and Greenland climate between 40 and 8 kyr BP. Nature, 366: 443-445.
Charles, C.D., and Fairbanks, R.G., 1992. Evidence from Southern Ocean sediments for the effect of North Atlantic deep-water flux on climate. Nature, 355: 416-419.
Clube, S.V.M. and Napier, W.M., 1984. The microstructure of terrestrial catastrophism. Mon. Not. R. Astr. Soc., 211: 953-968.
Coetzee, J.A., 1967. Pollen analytical studies in East and Southern Africa. Palaeoecology of Africa 3: 1-146.
Corliss, B.H., 1982. Linkage of North Atlantic and Southern Ocean deep-water circulation during glacial intervals. Nature, 298: 458-460.
Dansgaard, W., Clausen, H.B., Gundestrup, N., Hammer, C.U., Johnsen, S.F., Kristindottir, P.M., and Reeh, N., 1982. A new Greenland deep ice core. Science, 218: 1273-1277.
Dansgaard, W., White, J.W.C., and Johnsen, S.J., 1989. The abrupt termination of the Younger Dryas climate event. Nature, 339: 532-534.
Davelaar, J., Bleeker, J.M., Deerenberg, A.M. 1980. X-ray characteristics of Loop I and the local interstellar medium. Astron. Astrophys. 92: 231-237.
Denton, G., and Handy, C. H., 1994. Younger Dryas age advance of Franz Josef Glacier in the Southern Alps of New Zealand. Science, 264: 1434-1437.
Dingus, B.L. et al., 1988. Ultrahigh-energy pulsed emission from Hercules X-1 with anomalous air-shower muon production. Phys. Rev. Lett. 61: 1906-1909.
Dreimanis, A., 1966. The Susaca-interstadial and the subdivision of the late-glacial. Geol. en Mijnb., 45: 445-448.
Dreimanis, A. and Goldthwait, R.P., 1973. Wisconsin glaciation in the Huron, Erie, and Ontario lobes. In: R.F. Black and R.P. Goldthwait (Editors), The Wisconsin Stage (GSA Memoir 136). GSA, Boulder, CO, pp. 71-106.
Duplessy, J.C., Bé, A.W.H. and Blanc, P.L., 1981. Oxygen and carbon isotopic composition and biogeographic distribution of planktonic foraminifera in the Indian Ocean. Paleogeogr. Paleoclimatol. Paleoecol., 33: 9-46.
Emiliani, C., Rooth, C. and Stipp, J.J., 1978. The Late Wisconsin flood into the Gulf of Mexico. Earth Planet. Sci. Lett., 41: 159-162.
Fairbanks, R., 1989. A 17,000-year glacio-eustatic sea level record: influence of glacial melting rates on the Younger Dryas event and deep-ocean circulation. Nature, 342: 637-642.
Flower, B.P. and Kennett, J.P., 1990. The Younger Dryas cool episode in the Gulf of Mexico. Palaeocean. 5: 949-961.
Frisch, P.C., 1981. The nearby interstellar medium. Nature, 293: 377-379.
Frisch, P.C. and York, D.G., 1983. Synthesis maps of ultraviolet observations of neutral interstellar gas. Ap.J., 271: L59-L63.
Fuji, N., 1982. Paleolominological study of lagoon Kahoku-gata, Central Japan. XI INQUA Congress, Moscow, Vol. 1, p. 97.

Genthon, C. et al., 1987. Vostok ice core: climatic response to $CO_2$ and orbital forcing changes over the last climatic cycle. Nature, 329: 414-418.

Gold, T., 1969. Apollo II observations of a remarkable glazing phenomenon on the lunar surface. Science, 202, 1345-1347.

Grootes, P.M., 1978. Science, 200: 11.

Grün et al., 1993. Discovery of jovian dust streams and interstellar grains by the Ulysses spacecraft. Nature, 362: 428-430.

Harvey, L.D.D., 1980. Solar variability as a contributing factor to Holocene climatic change. Prog. Phys. Geog., 4: 487-530.

Hauser, M.G. et al., 1984. IRAS observations of the diffuse infrared background. Ap.J., 278: L15-18.

Heiles, C., Chu, Y.H., Reynolds, R.J., Yegingil, I., and Troland, T.H., 1980. A new look at the North Polar Spur. Ap.J., 242: 533-540.

Heusser, C.J., 1984. Late-glacial-Holocene climate of the lake district of Chile. Quat. Res., 22: 77-90.

Heusser, C.J. and Rabassa, J. 1987. Cold climatic episode of Younger Dryas age in Tierra del Fuego. Nature, 328: 609-611.

Heusser, C.J. and Streeter, S.S. 1980. A temperature and precipitation record of the past 16,000 years in Southern Chile. Science, 210: 1345-1347.

Horgan, J., 1995. Beyond Neptune, Scientific American 273 (10), pp. 24–26.

Hoyle, F. and Lyttleton, R.A., 1950. Variations in solar radiation and the cause of ice ages. J. Glaciol., 1: 453-455.

Hughen, K., et al., 1998. Deglacial changes in ocean circulation from an extended radiocarbon calibration. Nature, 391: 65 – 68.

Hughes, D. W., 1975. Cosmic dust influx to the Earth. Space Research, 15: 34.

Hyder, C.L., 1968. The infall-impact mechanism and solar flares. In: Y. Ohman (Editor), Mass Motions in Solar Flares and Related Phenomena. Wiley Interscience, New York, p. 57.

Ivy-Ochs, S., Schlüchter, C., Kubik, P. W., and Denton, G. H., 1999. Moraine exposure dates imply synchronous Younger Dryas glacier advance in the European Alps and in the Southern Alps of New Zealand. Geografiska Annaler., 81A: 313-323.

Jansen, E., and Veum, T., 1990. Evidence for two-step deglaciation and its impact on North Atlantic deep-water circulation. Nature, 343: 612-616.

Johnson, R.G. and Andrews, J.T., 1986. Glacial transitions in the oxygen isotope record of deep sea cores: Hypotheses of massive Antarctic ice-shelf destruction. Paleogeogr. Palaeoclimatol. Palaeoecol., 53: 107-138.

Johnsen, S.J. et al., 1992. Irregular glacial interstadials recorded in a new Greenland ice core. Nature, 359: 311-313.

Jouzel, J. et al. 1987. Vostok ice core: a continuous isotope temperature record over the last climatic cycle (160,000 years). Nature, 329: 403-407.

Konstantinov, A.N., Kocharov, G.E., and Levchenko, V.A., 1990. Explosion of a supernova 35,000 kyr ago. Soviet Astronomy Letters, 16:.343

Karrow, P.F., 1984. Quaternary stratigraphy and history, Great Lakes-St. Lawrence region. In: R.J. Fulton (Editor) Quaternary Stratigraphy of Canada (Geol. Survey Canada Paper 84-10). GSC, pp. 137-153.

Kelsall, T. et al., 1998. The COBE Diffuse Infrared Background Experiment search for the cosmic infrared background. II. Model of the interplanetary dust cloud. The Astrophysical Journal, 508: 44 – 73.

Kennett, J.P. and Shackleton, N.J., 1975. Laurentide ice sheet meltwater recorded in Gulf of Mexico deep-sea cores. Science, 188: 147-150.

Kudrass, H.R., Erienkeuser, H., Vollbrecht, R. and Weiss, W., 1991. Global nature of the Younger Dryas cooling event inferred from oxygen isotope data from Sulu Sea cores. Nature, 349: 406-409.

Lamb, R.C. et al., 1988. Tev gamma rays from Hercules X-1 pulsed at an anomalous frequency. Ap.J., 328: L13-L16.

Landgraf, M., et al., 2002. Origins of solar system dust beyond Jupiter. AJ, 123: 2857-2861.

LaViolette, P.A., 1983a. Galactic Explosions, Cosmic Dust Invasions and Climatic Change, Ph.D. dissertation, Portland State University, Portland, Oregon, 763 pp.

LaViolette, P.A., 1983b. Elevated concentrations of cosmic dust in Wisconsin Stage polar ice. Meteoritics 18: 336-337.

LaViolette, P.A., 1985. Evidence of high cosmic dust concentrations in Late Pleistocene polar ice (20,000 - 14,000 Years B.P.). Meteoritics, 20: 545-558.

LaViolette, P.A., 1987a. Cosmic-ray volleys from the Galactic Center and their recent impact on the Earth environment. Earth Moon Planets, 37: 241-286.

LaViolette, P. A., 2003. Galactic superwaves and their impact on the Earth. Starlane Publications, Niskayuna, NY.

LaViolette, P.A., 2005. Solar Cycle Variations in Ice Acidity at the End of the Last Ice Age: Possible Marker of a Climatically Significant Interstellar Dust Incursion. Planetary Space Science, 53: 385 - 393.

Ledru, M.P., 1993. Late Quaternary environmental and climatic changes in central Brazil. Quat. Res., 39: 90-98.

Lehman, S.J. and Keigwin, L.D., 1992. Sudden changes in North Atlantic circulation during the last deglaciation. Nature, 356: 757-762.

Leventer, A., Williams, D.F. and Kennett, J.P., 1982. Dynamics of the Laurentide ice sheet during the last deglaciation: evidence from the Gulf of Mexico. Earth Planet. Sci. Lett., 59: 11-17.

Leventer, A., Williams, D.F. and Kennett, J.P., 1983. Relationships between anoxia, glacial meltwater and microfossil preservation in the Orca Basin, Gulf of Mexico. Marine Geology, 53: 23-40.

Manabe, S. and Broccoli, A.J., 1985. The influence of continental ice sheets on the climate of an ice age. J. Geophys. Res., 90: 2167-2190.

McCrea, W., 1975. Ice ages and the Galaxy. Nature, 255: 607-609.

McCall, B. et al., 2003. An enhanced cosmic-ray flux towards ζ Persei inferred from a laboratory study of the $H_3^+$-$e^-$ recombination rate. Nature, 422: 500-502.

Mercer, J.H. and Palacios, O., 1977. Radiocarbon dating of the last glaciation in Peru. Geology, 5: 600-604.

Moore, P.D., 1981. Late glacial climatic changes. Nature, 291: 380.

Mörner, N.-A., 1973. Climatic changes during the last 35,000 years as indicated by land, sea, and air data. Boreas, 2: 33-52.

Mulvaney, R., et al., 2000. The transition from the last glacial period in inland and near-coastal Antarctica. Geophysical Research Letters, 27: 2673–2676.

Neftel, A., Oeschger, H., Staffelbach, T. and Stauffer, B., 1988. $CO_2$ record in the Byrd ice core 50,000 - 5,000 years BP. Nature, 331: 609-611.

North, G.R. and Crowley, T.J., 1985. Application of a seasonal climate model to cenozoic glaciation. J. Geol. Soc. (London), 142: 475-482.

Raisbeck, G.M., Yiou, F., Bourles, D., Lorius, C., Jouzel, J. and Barkov, N.I., 1987. Evidence for two intervals of enhanced $^{10}$Be deposition in Antarctic ice during the last glacial period. Nature, 326: 273-277.

Raisbeck, G.M. et al., 1981. Cosmogenic $^{10}Be$ concentrations in Antarctic ice during the past 30,000 years. Nature, 292: 825-826.

Raukas, A.V. and Serebryanny, L.R., 1972. On the Late Pleistocene chronology of the Russian platform, with special reference to continental glaciation. In: Proceedings 24th Intl. Geological Congress. Montreal, Quebec 1972, pp. 97-102.

Raynaud, D., Jouzel, J., Barnola, J.M., Chappellaz, J., Delmas, R.J., and Lorius, C., 1993. The ice record of greenhouse gases. Science, 259: 926-934.

Resvanis, L.K. et al., 1988. VHE gamma rays from Hercules X-1. Ap.J., 328: L9-L12.

Ruddiman, W.F., Sancetta, C.D. and McIntyre, A., 1977. Glacial/interglacial response rate of subpolar North Atlantic waters to climatic change, the record in oceanic sediments. Phil. Trans. R. Soc. Lond. B, 280: 119-142.

Ruddiman, W.F. and McIntyre, A., 1981. The North Atlantic ocean during the last deglaciation. Paleogeogr. Paleoclimatol. Paleoecol., 35: 145-214.

Schmidt, T. and Elasser, H., 1967. In: J.L. Weinberg (Editor) The Zodiacal Light and the Interplanetary Medium (SP-150). NASA, Washington, D.C., p. 301.

Schreve-Brinkman, E.J., 1978. A palynological study of the upper Quaternary sequence in the El Abra corridor and rock shelters (Colombia). Paleogeogr. Paleoclimatol. Paleoecol.,25: 1-109.

Schwarzschild, B., 1988. Are the ultra-energetic cosmic gammas really photons? Physics Today, 41(11): 17-23.

Scott, L., 1982. A Late Quaternary pollen record from the Transvaal Bushveld, South Africa. Quat. Res. 17: 339-370.

Sonett, C.P., 1991. A local supernova model shock ensemble using Antarctic Vostok ice core $^{10}Be$ radioactivity. December 1991 American Geophysical Union meeting, abstract in Eos 72: 72.

Steig, E. J. et al., 1998. Synchronous climate changes in Antarctica and the North Atlantic. Science 282: 92-95.

Stommel, H., 1980. Asymmetry of interoceanic fresh-water and heat fluxes. Proc. Natl. Acad. Sci. U.S.A., Geophys., 77(5): 2377-2381.

Sundquist, E.T., 1987. Ice core links $CO_2$ to climate. Nature, 329: 389.

Takahashi, H., Yokoyama, Y., Fireman, E.L., and Lorius, C., 1978. Iridium content of polar ice and accretion rate of cosmic matter. LPS 9: 1131.

Tauber, H., 1970. The Scandinavian varve chronology and $^{14}C$ dating. In: I. Olsson (Editor), Radiocarbon Variations and Absolute Chronology, Nobel Symp. 12. John Wiley & Sons, New York, pp. 179-196.

Taylor, K.C., Lamorey, G.W., Doyle, G.A., Alley, R.B., Grootes, P.M., Mayewskii, P.A., White, J.W.C. and Barlow, L.K., 1993. The 'flickering switch' of late Pleistocene climate change. Nature, 361: 432-436.

Van Campo, E., 1986. Monsoon fluctuations in two 20,000-Yr B.P. oxygen-isotope pollen records off southwest India. Quat. Res., 26: 376-388.

Van der Hammen, T., 1978. Stratigraphy and environments of the upper Quaternary of the El Abra corridor and rock shelters (Colombia). Paleogeogr. Paleoclimatol. Paleoecol., 25: 111-162.

Van der Hammen, T., Barfelds, J., de Jong, H. and de Veer, A.A., 1981. Glacial sequence and environmental history in the Sierra Nevada Del Cocuy (Columbia). Paleogeogr. Paleoclimatol. Paleoecol., 32: 247-340.

Veum, T., Jansen, E., Arnold, M., Beyer, I., and Duplessy, J.-C., 1992. Water mass exchange between the North Atlantic and the Norwegian Sea during the past 28,000 years. Nature, 356: 783-785.

Witte, M., Rosenbauer, H., Banaszkiewicz, M., Fahr, H., 1993. The ULYSSES neutral gas experiment - Determination of the velocity and temperature of the interstellar neutral helium. Advances in Space Research, 13(6): 121-130.

Woillard, G.M., and Mook, W.G., 1982. Carbon-14 dates at Grande Pile: Correlation of land and sea chronologies. Science, 215: 159-161.

Wright, H.E., 1984. Late glacial and late Holocene moraines in the Cerros Cuchpanga, Central Peru. Quat. Res. 21: 275-285.

Zook, H.A., Hartung, J.B. and Storzer, D., 1977. Solar flare activity: evidence for large-scale changes in the past. Icarus, 32: 106-126.

Zook, H. A., 1980. On lunar evidence for a possible large increase in solar flare activity ~2 X $10^4$ years ago. In R. Peppin, J. Eddy, and R. Merrill (eds.) Proceedings Conference on the Ancient Sun.